\documentclass{nature}
\usepackage{amsmath}
\usepackage[T1]{fontenc}
\usepackage{graphicx}
\usepackage{amssymb}
\usepackage{dcolumn}
\usepackage{color}
\usepackage{bm}
\usepackage[normalem]{ulem}
\usepackage{url}

\renewcommand{\vec}[1]{{\boldsymbol #1}}
\usepackage{graphicx}
\usepackage{pdfpages}

\begin{document}

\title{Electrical noise spectroscopy of magnons in a quantum Hall ferromagnet}


\author{Ravi Kumar$^{1}$\footnote{equally contributed}, Saurabh Kumar Srivastav$^{1}$\footnote{equally contributed}, Ujjal Roy$^{1}$\footnote{equally contributed}, Jinhong Park$^{2,3}$\footnote{equally contributed}, Christian Sp{\r a}nsl{\"a}tt$^{4}$, K. Watanabe$^{5}$, T. Taniguchi$^{5}$, Yuval Gefen$^{6}$, Alexander D. Mirlin$^{2,3}$, and Anindya Das$^{1}$\footnote{anindya@iisc.ac.in}}

\maketitle

\begin{affiliations}

\item Department of Physics, Indian Institute of Science, Bangalore, 560012, India.
\item Institute for Quantum Materials and Technologies, Karlsruhe Institute of Technology, 76021 Karlsruhe, Germany.
\item Institut f{\"u}r Theorie der Kondensierten Materie, Karlsruhe Institute of Technology, 76128 Karlsruhe, Germany.
\item Department of Microtechnology and Nanoscience (MC2), Chalmers University of Technology, S-412 96 G\"oteborg, Sweden.
\item National Institute of Material Science, 1-1 Namiki, Tsukuba 305-0044, Japan.
\item Department of Condensed Matter Physics, Weizmann Institute of Science, Rehovot 76100, Israel.

\end{affiliations}

\noindent\textbf{Collective spin-wave excitations -- magnons --  in a quantum Hall ferromagnet are promising quasi-particles for next-generation spintronics devices, including platforms for information transfer. Detection of these charge-neutral excitations relies on the conversion of magnons into electrical signals in the form of excess electrons and holes, but if these signals are equal the magnon detection remains elusive. In this work, we overcome this shortcoming by measuring the electrical noise generated by magnons. We use the symmetry-broken quantum Hall ferromagnet of the zeroth Landau level in graphene to launch magnons. Absorption of these magnons creates excess noise above the Zeeman energy and remains finite even when the average electrical signal 
is zero. Moreover, we formulate a theoretical model in which the noise is generated by equilibration (partial or full, depending on the bias voltage) between edge channels and propagating magnons. Our model, which agrees with experimental observations, also allows us to pinpoint the regime of ballistic magnon transport in our device.}


\noindent\textbf{Introduction.} The emergence of charge-neutral collective excitations presents a powerful platform for developing data processing as well as information transfer with small power consumption. Among these excitations, spin-wave excitations, or their quanta `magnons', in magnetic materials are promising. An obvious important task is to develop new techniques for the detection of these charge-neutral quasi-particles. So far, various experimental tools, such as inelastic neutron scattering\cite{yuan2020dirac,chen2021magnetic}, inelastic tunnelling spectroscopy\cite{ganguli2023visualization,spinelli2014imaging}, terahertz spectroscopy\cite{kampfrath2011coherent,zhuang2022excitation}, microwave Brillouin light scattering\cite{cramer2018magnon,wang2023observation}, nitrogen-vacancy centre\cite{carmiggelt2023broadband,lee2020nanoscale}, and superconducting qubits\cite{lachance2020entanglement} have been used to detect magnons in bulk magnetic materials. However, their detection in device geometries, which is necessary for information processing applications, has remained challenging until very recently. In particular, it was demonstrated by Wei et al.\cite{wei2018electrical} that magnons can be converted into electrical signals in a quantum Hall ferromagnet (QHF) in graphene.

Graphene offers a very versatile platform for new kinds of electronic devices. When subjected to a perpendicular magnetic field, graphene shows several unique quantum Hall (QH) phases, related to its peculiar sequence of Landau levels (LL), manifesting both spin and valley degrees of freedom\cite{novoselov2006unconventional,zhang2005experimental,gusynin2005unconventional}. In particular, the particle-hole symmetric zeroth LL (ZLL) has a rich variety of QHF phases\cite{yang2010hierarchy,abanin2006spin,weitz2010broken,zhao2010symmetry,sodemann2014broken,feldman2009broken,maher2014tunable}: When the ZLL is partially filled, Coulomb interactions break spin and valley symmetries, 
and for a quarter ($\nu = -1$) or three quarters ($\nu = 1$) filling, the QH phases comprise ferromagnetic insulator bulks with spin-polarized edge states\cite{yang1994quantum,young2012spin,kim2021edge,young2014tunable}. While the charge excitations in the bulk of these QHF insulators have a gap determined by the exchange energy ($E_{X} \sim \frac{e^2}{\epsilon \ell_{B}}$, where $e$, $\epsilon$ and $\ell_B$ are the elementary charge, dielectric constant, and the magnetic length), the spin-waves (magnons) have instead a gap determined by the Zeeman energy ($E_{Z} = g\mu_{B}B$, where $g$ is the Land\'e g-factor, and $\mu_B$ is the Bohr magneton)\cite{girvin} and are in fact the lowest energy excitations of the system. However,  magnons do not carry electrical charge, and therefore do not have a large impact on electrical transport, which in turn makes it a difficult task to detect them. There are a few reported attempts of generating and detecting spin-wave excitations or
magnons in graphene-based QHF  devices\cite{wei2018electrical,Stepanov2018,Zhou2020,assouline2021excitonic,fu2021gapless,pierce2022thermodynamics}. While magnon generation in these phases is based on an out-of-equilibrium occupation of edge channels with opposite spin, the detection of the magnons relies on the absorption of magnons by edge modes in the vicinity of ohmic contacts. The absorption of magnons by the edge modes creates excess electrons or holes in different corners of the graphene devices, and the measured electrical signal depends on the relative difference between the electron and hole signal magnitudes, which, in turn, critically depend on the device geometry. One may, therefore, not be able to detect any electrical signal if both the excited electrons and hole signals are equal. Thus, an alternative technique, which does not rely on the difference between excess electron and hole signals, is necessary for sensitive detection of magnons. 

In this work, we demonstrate 
that electrical noise spectroscopy of magnons is a powerful method that satisfies the detection sensitivity requirement. 
We first establish that our device hosts symmetry-broken robust QH phases and study the magnon transport when the bulk filling is kept at $\nu = 1$. In order to generate magnons, we inject an edge current through an ohmic contact. While the injected current only flows in the downstream direction (as dictated by the electron motion subject to external magnetic field), we measure the non-local electrochemical potential of a floating ohmic contact placed upstream from the source contact. Whenever the bias voltage of the injection contact corresponds to an energy smaller than the Zeeman energy $E_{Z}$, no non-local signal is detectable. 
As the bias energy exceeds $E_{Z}$, we measure a finite non-local signal for negative bias voltages.
By contrast, the non-local signal remains zero for the entire positive bias voltages, which may naively suggest that magnons are not generated in this bias regime. Next, we switch to measuring the electrical noise and show that, as expected, no noise is detected below $E_{Z}$. 
On the other hand, as soon as the bias energy exceeds $E_{Z}$, the noise increases for both signs of the bias voltage. 
We show that the noise contributions created due to magnon absorption at different corners in our devices are additive, even when the average electron and hole currents mutually cancel (which happens for positive bias voltages). This renders noise spectroscopy a highly sensitive tool for magnon detection. Finally, our theoretically calculated noise captures well the experimental data and further suggests that the detected noise is a result of an increase of the effective temperature of the system as a result of equilibration between edge channels and magnons. 

\begin{figure}
\begin{center}
\centerline{\includegraphics[width=1.0\textwidth]{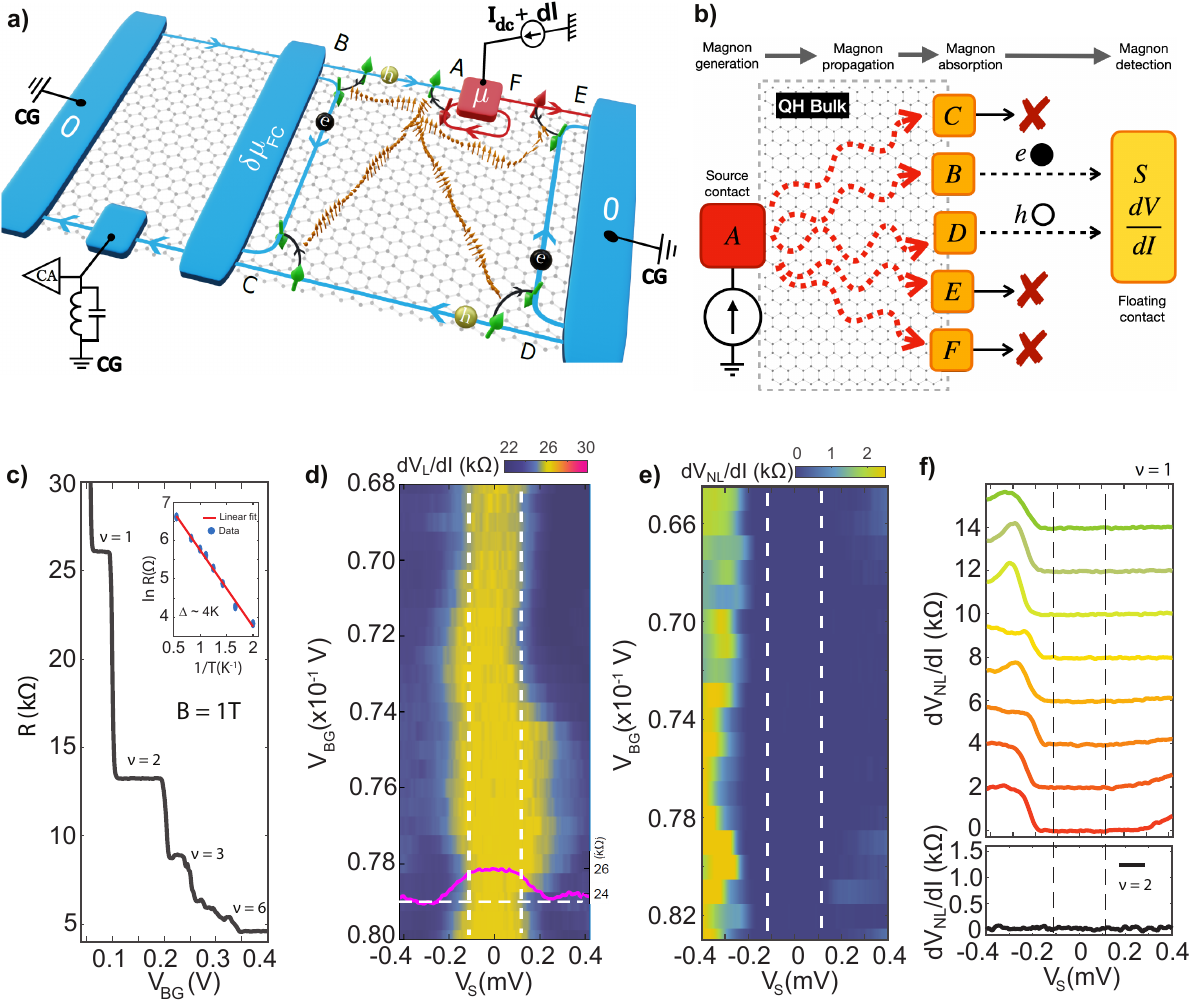}}
\caption{\textbf{Device schematic, magnon generation and detection in quantum Hall ferromagnet.} (\textbf{a}) 
The device has a left, right, transverse, and a floating contact. The device is set to $\nu = 1$, whereas regions adjacent to the contacts are tuned to $\nu = 2$, as shown by the additional circulating inner edges near the contacts. The spin polarization of the outer and inner edges are orthogonal, denoted by up and down arrows, respectively. A dc plus ac current ($I_{dc} + dI$) is injected through the upper red transverse contact, and when the electrochemical potential ($\mu$) exceeds Zeeman energy ($E_Z$), magnons are generated near point \textbf{'A'} via a spin-flip process. These magnons propagate through the QH bulk and are absorbed at other corners via the reverse spin-flip process. The bottom transverse contact is used to measure the voltage ($dV$) and noise ($S$) of the floating contact using standard lock-in ($\sim 13 Hz$) and LCR resonance circuit ($\sim 740 kHz$), respectively. (\textbf{b}) Magnon absorption at the different corners creates electron-hole excitations, but only points \textbf{'B'} and \textbf{'D'} contribute excess electrons and holes to the floating contact, respectively. (\textbf{c}) QH response at $B = 1$T. The inset shows the activation gap of $\nu = 1$, which is $\sim 4$K. (\textbf{d}) 2D colour map of the differential resistance ($dV_L/dI$) measured at the source contact vs the dc bias voltage ($V_{S} = I_{dc} \times \frac{h}{e^2}$) and the gate voltage around the centre of the $\nu = 1$ plateau. A line cut at $V_{BG} = 0.079$ is shown in solid magenta. (\textbf{e}) Non-local $dV_{NL}/dI$ of the floating contact vs source and gate voltages. (\textbf{f}) (upper panel) Line cuts from \textbf{e)}. Each plot is shifted vertically for clarity. (bottom panel) Non-local $dV_{NL}/dI$ for bulk $\nu = 2$. The vertical lines in d), e), and f) represent the Zeeman energy at B = 1T.}
\label{Figure1}
\end{center}
\end{figure}

\noindent\textbf{Device and experimental principle.}
Figure~1(a) shows the schematics of our device and measurement setup. The device consists of $hBN$ encapsulated graphite-gated high-mobility single-layer graphene, fabricated by the standard dry transfer technique\cite{purdie2018cleaning,pizzocchero2016hot}. Device fabrication and characterization are detailed in the Supplementary Information (SI-S1). The QH response of the device at a magnetic field ($B$) 
of $1$\,T is shown in Fig. 1(c), indicating robust QH plateaus and the inset depicts the activation gap at $\nu = 1$, which is estimated to be $\sim 4K$ (see SI-S1). 
As seen in Fig. 1(a), the device has left and right ground contacts, while the upper transverse contact is utilized to inject current for magnon generation. The lower transverse contact is used to detect the change in chemical potential of the floating contact (FC) due to magnon absorption. The device's bulk is tuned to the $\nu = 1$ QHF 
state, allowing it to host magnons. Importantly, the local doping due to the attached metallic contacts increases the filling factor to $\nu = 2$ near these contacts; this is represented (shown only for the right side of the FC) by additional loop-shaped edge modes at each contact, and are referred to 
as the ``inner edge''. In contrast, the outer edge propagates between contacts, as shown in Fig. 1(a). A noiseless current, $I_{dc} +dI$, comprising a dc and an ac component, is injected into the red-coloured source contact in Fig.~1(a). The injected current flows along the outer edge with up-spin polarization. This current exits the sample at the right-most grounded contact. The current along the inner edge, which flows around the source contact, has a down-spin polarization, 
does not contribute to the electrical conductance in the circuit. The dc voltage drop at the source contact, $V_{S} = I_{dc} \times \frac{h}{e^2}$, is shown as the electrochemical potential $\mu$ in Fig.~1(a). The corresponding ac voltage that drops at the source contact is $dV_{L} = dI \times \frac{h}{e^2}$. Whenever $\mu$ exceeds the Zeeman energy $E_Z$, i.e., $\vert \mu \rvert \geq E_Z$, the electrons flowing along the circulating inner edge can tunnel into the outer edge by flipping their spin via magnon emission near point \textbf{`A'}, as shown in Fig.~1(a). This process does not directly alter the electrical conductance since the tunnelling current flows back and is absorbed by the same injection contact. The emitted magnons propagate through the bulk of the device and can be absorbed at the device corners (\textbf{`B'}, \textbf{`C'}, \textbf{`D'}, \textbf{`E'} and \textbf{`F'}) via tunneling of electrons from the outer edge to the inner edge through the reverse spin flipping process. 
However, only parts of the currents generated at the two corners \textbf{`B'} and \textbf{`D'} arrive at the FC and contribute to the fluctuations of the electrochemical potential $\delta \mu_{\text{FC}}$ of the FC. This is so since generated electron and hole excitations are separated at points \textbf{`B'} and \textbf{`D'} into two respective currents, only one of which flows towards the FC as shown in Figs.~1(a) and 1(b). The fluctuations $\delta \mu_{\text{FC}}$ 
are measured in the lower transverse contact placed to the left of the FC on the lower edge, by using an LCR resonance circuit at a frequency of 
$\sim740 kHz$, followed by an amplifier chain and a spectrum analyzer (see Ref.~\citeonline{doi:10.1126/sciadv.aaw5798,Ravi2022}, Methods, and SI-S8). We also measure the average chemical potential of the FC ($dV_{NL}$) via the same transverse contacts with standard lock-in measurements. It should be noted that the magnon generation in Fig.~1(a) is shown only for negative bias voltage; for positive bias voltage, magnons are instead generated near point \textbf{`E'}, as shown in Fig.~2(b). 
We carried out measurements in two devices, where for the second device (bilayer graphene), the filling near the contacts was tuned by local gating, showing similar results (see SI-S6 and S7).

\noindent\textbf{Magnon detection using non-local resistance and noise spectroscopy.}
Figure~1(d) shows a 2D color map of the differential resistance $R_{L} = dV_{L}/dI$ (with $L$ denoting  ``local'') measured in the injection contact as a function of the bias voltage ($V_{S}$) and gate voltage ($V_{BG}$) around the center of the $\nu = 1$ plateau. It can be seen that within $E_Z$ 
[white vertical dashed lines in Fig.~1(d)], $R_{L}$ remains constant at $\frac{h}{e^2} \sim 25.8 k\Omega$ and decreases on both sides above $E_Z$, as shown by the solid magenta line in Fig.~1(d). This feature is similar to that in Ref.~\citeonline{wei2018electrical}, and can be understood as follows: For negative bias voltages, magnons are generated at \textbf{`A'}. Absorption at \textbf{`B'} and \textbf{`F'} reduce (via holes) and increase (via electrons) the chemical potential ($dV_{L}$) of the source contact, respectively, and thus affect $R_{L}$. However, since the absorption at \textbf{`B'} dominates over that at \textbf{`F'}, $R_{L}$ decreases. Note that to be absorbed at \textbf{`F'}, the magnons have to bend around the injected contact in contrast to their straight propagation when reaching \textbf{`B'}. Similarly, for positive bias voltage, the generated magnons from \textbf{`E'} [see Fig.~2(b)] are absorbed dominantly at \textbf{`F'} in comparison to \textbf{`B'} and thus $R_{L}$ decreases.

A more powerful approach to magnon detection, which permits to explicitly demonstrate and to explore magnon transport through the system, is provided by non-local measurements\cite{wei2018electrical,fu2021gapless}.
Figure~1(e) shows a 2D color map of the non-local differential resistance, $R_{NL} = dV_{NL}/dI$, vs bias and gate voltages, where $dV_{NL}$ is the chemical potential of the FC. As seen from the line cut in Fig. 1(f) (top panel), $R_{NL}$ remains zero within $E_Z$ (vertical, dashed lines), and increases for negative bias voltage above $E_Z$. However, $R_{NL}$ is almost zero for the entire positive bias voltage range. When the bulk filling was set to $\nu = 2$, no detectable non-local signal (Fig. 1(f), lower panel) was observed as the ground state is then non-magnetic. The $R_{NL}$ in Figs. 2(e) and 2(f) can be understood as follows: As schematically shown in Fig. 1(b), the magnon absorption at \textbf{`B'} and \textbf{`D'} contributes to the non-local signal of the FC via excess electrons and holes, respectively. For negative bias voltage, the magnons are generated at \textbf{`A'}, but the absorption at \textbf{`B'} dominates over \textbf{`D'} due to shorter distance [Fig. 1(a)], and thus $R_{NL}$ takes a finite value. However, for positive bias voltage, the magnons are generated at \textbf{`E'} [Fig. 2(b)], and the absorption at \textbf{`B'} and \textbf{`D'} are almost equal due to their similar distance from \textbf{`E'}. Thus, $R_{NL}$ becomes almost zero. 

\begin{figure}
\begin{center}
\centerline{\includegraphics[width=1.0\textwidth]{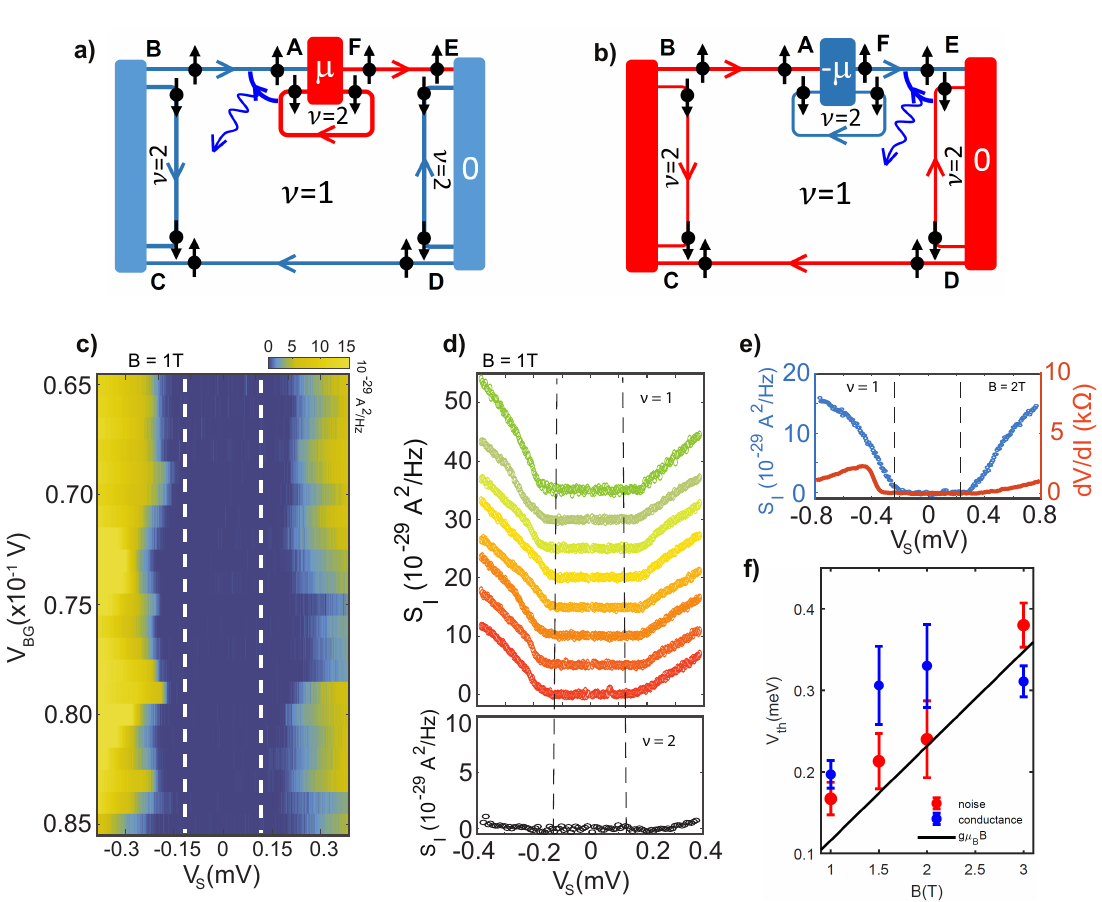}}
\caption{\textbf{Noise spectroscopy of magnons.} Magnon generation for negative (\textbf{a}) and positive (\textbf{b}) bias voltages, where magnons are generated at point \textbf{'A'} and \textbf{'E'}, respectively. The generated magnons propagate through the QH bulk and are absorbed at different corners. Only magnon absorption at points \textbf{'B'} and \textbf{'D'} generates noise at the floating contact. (\textbf{c}) 2D colour map of excess noise generated at the floating contact for different bias and gate voltages. (\textbf{d}) (top panel) Line cuts of excess noise from c) for different $V_{BG}$ around the centre of the $\nu = 1$ plateau. Each plot is shifted vertically for clarity. (bottom panel) Noise spectra for bulk $\nu = 2$. As expected, no excess noise is visible. (\textbf{e}) Noise spectra and $dV/dI$ for bulk filling $\nu = 1$ at $B = 2$T. The vertically dashed lines in c), d), and e) depict the  Zeeman energy $E_Z$. (\textbf{f}) Thresholds of the bias voltage (with error bars) vs magnetic field. The thresholds are extracted both from noise spectroscopy (solid red circles) and non-local differential resistance measurements (solid blue circles). Plotted is also $E_Z = g \mu_B B$ (solid black line).
}
\label{Figure2}
\end{center}
\end{figure}

Figure~2(c) shows a 2D color map of the measured excess noise ($S_I$) in the FC as a function of bias and gate voltages. The corresponding line cuts are shown in Fig. 2(d) (upper panel). We see that that $S_I$ remains zero as long as $|eV_S| \le  E_Z$, and keeps increasing for larger values of either positive or negative bias voltage. 
This feature stands in stark contrast to Figs.~1(e) and 1(f). No significant noise was detected for the non-magnetic state at $\nu = 2$, shown in Fig. 2(d) (lower panel), which establishes that the contribution from other degrees of freedom in the bulk (e.g., phonons) are negligible in our experiment. We have repeated this measurement at different magnetic fields (see SI-S2). For example, Fig. 2(e) shows $R_{NL}$ and $S_I$ at $B = 2T$, which display features very similar as the data at $B = 1$T. The threshold voltage, $V_{\rm th}$ for magnon detection, extracted from $S_I$ at different $B$, is plotted in Fig.~2(f) (solid red circles) along with $E_{Z} = g \mu_B B$ (solid black line), and we see that the two are strongly correlated. We also show the threshold voltage extracted from the non-local resistance (for negative bias voltage) as solid blue circles. At a given magnetic field, the threshold voltage was extracted for several points across the plateau, and its mean value and standard deviation are shown in Fig. 2(f) (see SI-S3). The key insight from our noise spectroscopy of magnons is that although $R_{NL}$ is almost zero due to the competition between excess electrons at \textbf{`B'} and excess holes at \textbf{`D'}, the fluctuations (variance) are instead additive, making $S_{I}$ very sensitive for magnon detection without any significant device geometry dependence.

We further note that the threshold voltage above which the non-local resistance arises is significantly higher than $E_Z$ [see Figs.~2(e) and 2(f)] except for $B = 3$T. This behavior has been observed in previous works as well\cite{wei2018electrical, pierce2022thermodynamics}. In contrast to the resistance data, however, the noise starts to increase at bias voltage $|eV_S| \sim E_Z$ 
[see Figs.~2(d), 2(e), and 2(f)]. The difference in threshold voltages for the non-local resistance and the noise can be understood if magnons are absorbed at $ \textbf{`B'}$ and $\textbf{`D'}$ with equal probabilities within the bias voltage window $E_Z < |eV_{S}| < e V_{\text{th}}$. Hence, this absorption process is invisible in the non-local resistance data while strikingly visible in the noise data. Such an equal magnon absorption may arise from a ballistic magnon transport in the bias voltage window $E_Z < |eV_{S}| < e V_{\text{th}}$, where generated magnons propagate with a long wavelength  $\lambda \gg \ell_B$. Such magnons experience little scattering from other degrees of freedom, particularly phonons or skyrmions\cite{Zhou2020}. {However, the ballistic motion of magnons may not be possible at a higher magnetic field, $B = 3$T. At this magnetic field, the phonons play an important role since a larger current is required in order to generate magnons due to the higher $E_Z$. Thus, the increased dissipation near \textbf{`A'} and \textbf{`E'} [see Fig. 1(a)] is able to excite phonons (see SI-S5). As a result, the threshold voltage for the non-local resistance at $3$T is reduced to the vicinity of 
$E_Z$, and in fact, is even slightly lower than $E_Z$ due to the temperature broadening effect. A similar result was observed at elevated bath temperatures at $B = 1$T, as shown in SI-S4. 


\begin{figure}
\begin{center}
\centerline{\includegraphics[width=0.9\textwidth]{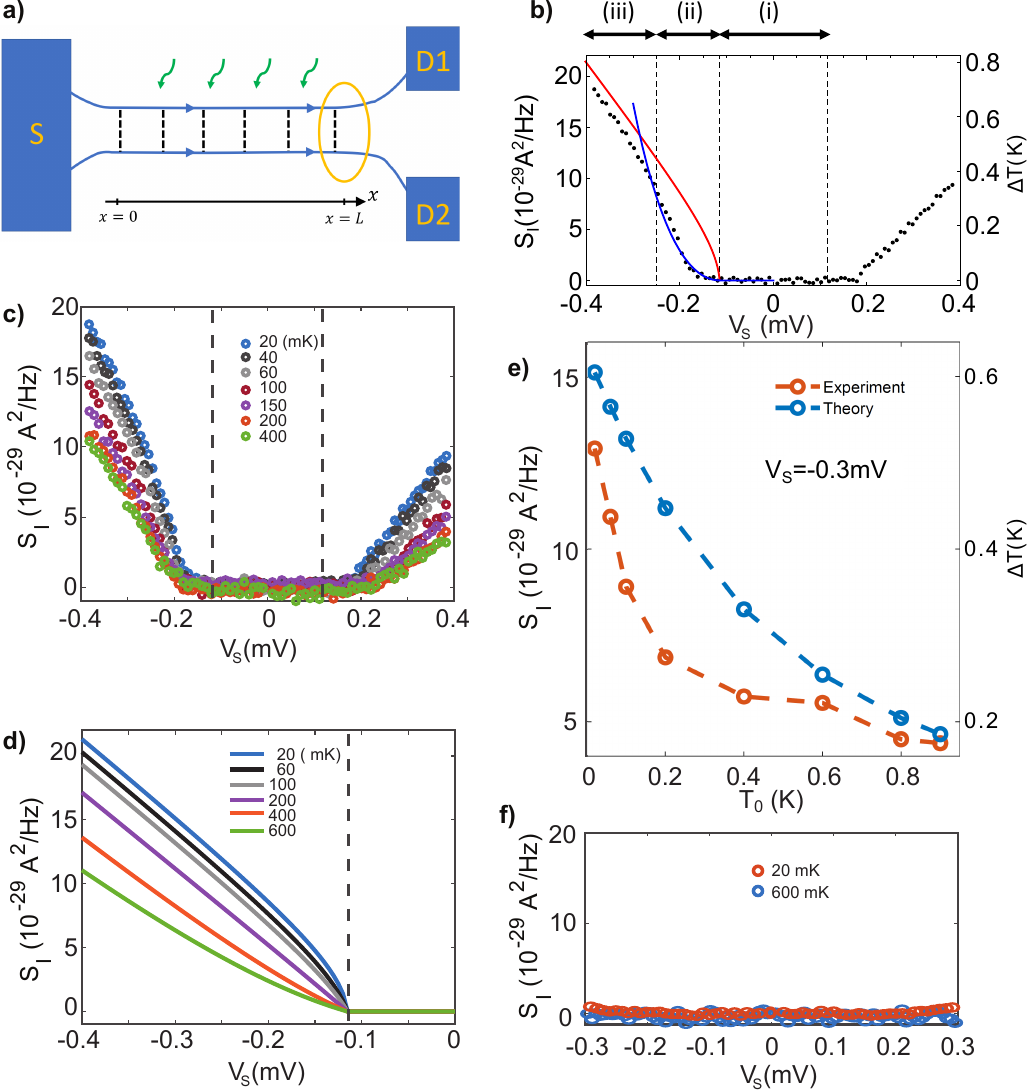}}
\caption{\textbf{Theoretical model and temperature dependence of excess noise.} (\textbf{a}) The magnon absorption (wiggly green lines with arrow) at any corner is modelled as a line segment of co-propagating edges, where tunnelling of electrons occur from outer to inner edge [Figs. 1(a), 2(a), and 2(b)]. The noise
from the total tunnelling current is dominantly generated in the vicinity of $x = L$ (yellow circle), where the local equilibrium noise dominates over the shot noise. (\textbf{b}) Comparison between the experimentally measured excess noise at $20$\,mK (black solid circles) with the theoretically calculated data: The red solid line is the prediction for the strongly equilibrated regime and the blue solid line is for the partially equilibrated regime. These regimes of no magnons, partial and strong equilibration are further indicated by the horizontal arrows at the top of the axis. The right-hand side axis indicates the excess temperature, see Eq.~\eqref{eq:noise} (\textbf{c}) Measured noise at different bath temperatures. (\textbf{d}) Noise calculated from Eqs.~\eqref{eq:noise}-\eqref{eq:T} at different bath temperatures $T_0$. (\textbf{e}) Comparison between experiment and theory for the excess noise at $V_{S} = -0.3$\,mV as a function of bath temperature. The right side of the axis indicates the excess temperature. (\textbf{f}) Excess noise for bulk filling $\nu=2$ at $20$\,mK and $600$\,mK, both at $B=1$T.}
\label{Figure3}
\end{center}
\end{figure}

\noindent\textbf{Theoretical model and comparison to experiment.}
We model the edge segments where the magnon generation and absorption take place as line junctions of 
co-propagating edges with length $L$, where electrons tunnel
between edge channels (with spin-$\uparrow$ and spin-$\downarrow$), see Fig.~3(a). 
Each such tunnelling event is associated with the generation or absorption of magnons. We identify two distinct transport regimes depending on a degree of equilibration, characterized by the equilibration length $\ell_{\text{eq}}$; a short-junction regime ($L < \ell_{\text{eq}}$) with partial equilibration of the edge channels and the magnons, and a long-junction regime ($L > \ell_{\text{eq}}$) with strong equilibration, see Methods and SI-S9 for details. In the strong equilibration regime, 
equilibration in the magnon-generation region \textbf{`A'} takes place until the chemical potential difference between the edge channels equals $E_Z$. At this saturation point, further magnon generation is strongly suppressed. All generated magnons propagate in the bulk of the QH state and are eventually absorbed in one of the absorption regions (\textbf{`B'}, \textbf{`C'}, \textbf{`D'}, \textbf{`E'} and \textbf{`F'}). Each absorption event creates an electron-hole pair (an electron in the spin-$\downarrow$ channel and a hole in the spin-$\uparrow$ channel). 
%
These pairs produce the measured excess noise. In each absorption line junction, the excess noise is dominantly generated near $x = L$ [yellow circle in Fig.~3(a)] while remaining contributions are exponentially suppressed, see  Refs.~\citeonline{Park2019,Spanslatt2019,Srivastav2019,Ravi2022,Srivastav2022} for a similar noise-generating mechanism. 
The excess noise $S_I$ reflects an increased temperature of the edge channels during the magnon absorption, given by
\begin{align} \label{eq:noise}
S_I = \frac{1}{2} \left(\frac{e^2}{h} (T_0 + T)  - 2  \frac{e^2}{h} T_0  \right)
 =   \frac{1}{2} \left(\frac{e^2}{h} ( T - T_0 ) \right)\,,
\end{align}
where $T_0$ is the bath temperature and
\begin{align}
    T = \sqrt{T_0^2 + \frac{3 (|eV_S| - E_Z) (2E_Z + 3|eV_S|)}{ (5 \pi)^2} \theta (|eV_S| - E_Z)} 
    \label{eq:T}
\end{align}
is the effective temperature of the system as a result of equilibration.
Furthermore, $\theta(|eV_S|-E_Z)$ is the step function, which reflects the fact that no magnons can be absorbed for bias energies below 
$E_Z$. The factor $1/2$ in Eq.~\eqref{eq:noise} originates from the noise-measurement scheme, see Methods. 
In Fig. 3(b), we compare our theoretically calculated excess noise (solid red line), $S_I$, with the experimentally measured noise versus the bias energy $eV_S$ (for simplicity, only the negative bias side is displayed), at fixed $T_0=20$mK. Figure~3(c) shows the measured noise at different bath temperatures ($T_0$), and the corresponding theoretical plots are shown in Fig. 3(d). A comparison between the experiment (orange circles) and theory (blue circles) for $S_I$ at $V_{S} = -0.3$\,mV as a function of $T_0$ is shown in Fig. 3(e). Our theoretical model captures well the characteristic features of the noise. 
Note that as seen in Fig.~3(f), no excess noise was detected even at higher temperatures ($600$mK) for $\nu = 2$ at $B = 1$T. 

The bias voltage dependence of the excess noise defines three regimes in Fig.~3(b); (i) Biases $|eV_S| < E_Z$ result in no magnon generation and thus no excess noise. (ii) In a narrow region $0 < |eV_S| - E_Z < \frac{1}{\gamma L}$, the equilibration in magnon absorption and generation regions is only partial, $L <\ell_{\textrm{eq}} \equiv \frac{1}{\gamma (|eV_S| - E_Z)}$. Here, $\gamma$ is a parameter proportional to the tunneling strength in every tunnel junction comprising the line junction. This lack of equilibration allows us to model the magnon-generation and absorption regions as single tunnel junctions in regime (ii). In such a model, the noise generation is of non-equilibrium nature, resulting in $S_I \propto (|eV_S| - E_Z)^2$, shown in Fig.~3(b) by the solid blue line. (iii) For larger biases $|eV_S| > E_Z + \frac{1}{\gamma L}$ and hence $\ell_{\textrm{eq}} <L$, the edge channels and magnons achieve full equilibration in the magnon absorption and generation regions. We find that our theoretical model is in good agreement with the experimental data. In particular, at sufficiently large biases [regime (iii)], our equilibrated line junction model correctly describes several experimental observations: the sudden increase followed by (approximate) saturation of the non-local conductance as a function of the bias voltage [see Fig. 1(f)], the linear behavior of the noise as a function of the bias voltage [Fig. 2(d)], and the temperature dependence of the excess noise [Fig. 3(d)-(e)]. In addition, our single tunnel junction model [partial equilibration regime (ii)] properly describes the crossover region of bias voltages close to $E_Z$. Note that our theory assumes that magnons are absorbed in all the absorption regions with the same probability, but in reality, there may be deviations. These can explain some variations between experimental data curves.  


\noindent\textbf{Discussion.} As we show in Eq.~\eqref{eq:noise}, the excess noise generated in the line junction [regime (iii)] reflects the increase in temperature $T-T_0$ of the edge due to heating. The temperature behavior extracted from the measured excess noise data in Fig. 3(b) (right y-axis) is similar to the temperature behavior in Fig. 3(d) of Ref.~\citeonline{pierce2022thermodynamics}, a result which was obtained from $R_{xx}$ thermometry measurements. However, we have never observed any saturation of the temperature at high bias voltage, which was argued to be indicative of strong equilibration between free magnons and skyrmions. 
Furthermore, the noise behavior as a function of the bias voltage appears to be correlated with that of the visibility of the Mach-Zender interferometry measured in Ref.~\citeonline{assouline2021excitonic}. It would be interesting to make a detailed connection between those two different quantities.
Finally, we emphasize that our measurements were performed for relatively small magnetic fields and lower ambient temperatures than in previous works\cite{wei2018electrical,assouline2021excitonic,pierce2022thermodynamics}. These small quantities allow us to fully neglect the effect of phonons. We have observed a sizeable effect of phonons only for magnetic fields $B > 2$T (see SI-S5). 

\noindent\textbf{Conclusion.}
We have demonstrated the utility of electrical noise spectroscopy as a highly sensitive tool for detecting and studying magnons in a quantum Hall ferromagnet. Our new protocol overcomes non-universal (device geometry dependent) features that screen out the presence of magnons, when other detection tools are employed, most prominently non-local conductance measurements. This robustness paves the way for utilizing magnons as low-power
information carriers in future quantum technologies. Intriguing generalizations of our approach, with a promise of novel physics, include bulk phases of 
the fractional quantum Hall regime as well as of integer and fractional Chern insulator phases of twisted bilayer graphene\cite{nuckolls_strongly_2020,das_symmetry-broken_2021,Xie2021}. Further implementations of our approach may include other ferromagnetic materials and vdW magnets\cite{Gong2017,Huang2017}.


\newpage
\newpage
\newpage
\newpage
\newpage
\newpage

\section*{Methods.}
\subsection{Device and measurements scheme.} Utilizing the dry transfer pick-up approach, we fabricated encapsulated devices consisting of a heterostructure involving hBN (hexagonal boron nitride), single-layer graphene (SLG), and graphite layers. The procedure for creating this heterostructure comprised the mechanical exfoliation of hBN and graphite crystals onto an oxidized silicon wafer through the widely employed scotch tape method. Initially, a layer of hBN, with a thickness of approximately 25-30 nm, was picked up at a temperature of 90°C. This was achieved using a Poly-Bisphenol-A-Carbonate (PC) coated Polydimethylsiloxane (PDMS) stamp on a glass slide attached to a home-built micromanipulator. The hBN flake was aligned over the previously exfoliated SLG layer picked up at 90°C. The subsequent step involved picking up the bottom hBN layer of similar thickness. Following the same process, this bottom hBN was picked up utilizing the previously acquired hBN/SLG assembly. After this, the hBN/SLG/hBN heterostructure was employed to pick up the graphite flake. Ultimately, this resulting heterostructure (hBN/SLG/hBN/graphite) was placed on top of a 285 nm thick oxidized silicon wafer at a temperature of 180°C. To remove the residues of PC, this final stack was cleaned in chloroform (CHCl3) overnight, followed by cleaning in acetone and isopropyl alcohol (IPA). After this, Poly-methyl-methacrylate (PMMA) photoresist was coated on this heterostructure to define the contact regions using electron beam lithography (EBL). Apart from the conventional contacts, we defined a region of $\sim$ $6 \mu m^2$ area in the middle of SLG flake, which acts as a floating metallic reservoir upon edge contact metallization. After EBL, reactive ion etching (mixture of CHF3 and O2 gas with a flow rate of 40 sccm and 4 sccm, respectively, at 25°C with RF power of 60W) was used to define the edge contact. The etching time was optimized such that the bottom hBN did not etch completely to isolate the contacts from the bottom graphite flake, which was used as the back gate. Finally, the thermal deposition of Cr/Pd/Au (3/12/60 nm) was done in an evaporator chamber with a base pressure of $\sim$ $1 \times 10^{-7}$ mbar. After deposition, a lift-off procedure was performed in hot acetone and IPA. The device's schematics and measurement setup are shown in Fig. 1(a). The distance from the floating contact to the ground contacts was $\sim 5 \mu m$, whereas the transverse contacts were placed at a distance of $\sim 2.5 \mu m$. 

All measurements were done in a cryo-free dilution refrigerator with a $\sim$ 20 mK base temperature. The electrical conductance was measured using the standard lock-in technique, whereas the noise was measured using an LCR resonant circuit at resonance frequency $\sim$ 740 kHz. The signal was amplified by a homemade preamplifier at 4 K followed by a room temperature amplifier, and finally measured by a spectrum analyzer. At zero bias, the equilibrium voltage noise measured at the amplifier contact is given by
\begin{equation}
S_V = g^{2}(4k_{B}TR+V_{n}^{2}+i_{n}^{2}R^{2})BW \,,
\end{equation} 
where $k_{B}$ is the Boltzmann constant, $T$ is the temperature, $R$ is the resistance of the QH state, $g$ is the gain of the amplifier chain, and $BW$ is the bandwidth. The first term, $4k_{B}TR$, corresponds to the thermal noise, and $V_{n}^{2}$ and $i_{n}^{2}$ are the intrinsic voltage and current noise of the amplifier. At finite bias above the Zeeman energy, due to magnon absorption at points \textbf{`B'} and \textbf{`D'}, chemical potential fluctuations of FC create excess voltage noise at the amplifier contact. At the same time, the intrinsic noise of the amplifier remains unchanged. Due to the white nature of the thermal noise and the excess noise, we could operate at higher frequency ($\sim$740\,kHz), which eliminates the contribution from flicker noise (1/f) which usually becomes negligible for frequencies above few tens of Hz.  The excess noise ($\delta S_{V}$) due to bias current is obtained by subtracting the noise value at zero bias from the noise at finite bias, i.e $\delta S_{V}=S_{V}(I)-S_{V}(I=0)$. The excess voltage noise $\delta S_{V}$ is converted to excess current noise $S_{I}$ according to $S_{I}=\frac{\delta S_{V}}{R^2}$, where $R=\frac{h}{\nu e^2}$ is the resistance of the considered QH edge.

\subsection{Theoretical calculation of the non-local resistance and noise.}
\label{sec:MethodsTheory}
To compute the tunneling current, non-local resistance, and noise generated in the magnon absorption regions, we model the magnon generation and absorption regions as line junctions of length $L$. These line junctions are modelled as extended segments with two co-propagating edge channels in which electrons tunnel along a series of tunnel junctions, see Fig.~3(a). We identify two distinct transport regimes: those of a short (partially equilibrated; $L < \ell_{\text{eq}}$) and long (equilibrated; $L > \ell_{\text{eq}}$) junctions, where $\ell_{\text{eq}}$ is the equilibration length. 
The short-junction regime can be equivalently modelled as a single tunnel junction. 
Details of the theoretical analysis are presented in SI, Sec.~S9.


We first consider the partial-equilibration regime, treating the magnon generation or absorption regions as a single tunnel junction (at position $x=0$). The Hamiltonian describing this junction reads
\begin{align}
    H &= - i v \sum_{s = \uparrow, \downarrow} \int dx \psi_s^{\dagger} (x) \partial_x \psi_s(x) + \sum_{\vec{q}}(E_Z + \hbar \omega_{\vec{q}}) b_{\vec{q}}^{\dagger} b_{\vec{q}} \notag \\&+ W\psi_\uparrow^{\dagger} (x = 0) \psi_\downarrow (x = 0) b^{\dagger} (x = 0)  + \textrm{h.c.}.
    \label{method:H}
\end{align}
Employing the Keldysh non-equilibrium formalism, we derive zero-temperature expressions for the tunneling current $I_{\text{ab}}$, non-local resistance $dV_{\text{ab}}/dI$, and noise $S_{\text{ab}}$ generated in an absorption region, respectively: 
\begin{align}
    I_{\text{ab}} &= \gamma' \frac{e}{2h}  (|eV_S|- E_Z)^2 \theta(|eV_S| - E_Z)\,,
    \label{method:singlejunctioncurrent}\\
    \left |\frac{dV_{\text{ab}}}{dI} \right | & =  \gamma' \frac{h}{e^2} (|eV_S|- E_Z) \theta(|eV_S| - E_Z)\,,
    \label{method:singlejunctionnonlocal}\\  
    S_{\text{ab}} &=  \frac{e^2}{h} \gamma' (|eV_S| - E_Z)^2 \theta(|eV_S| - E_Z)\,. \label{method:singlejunctionnoise}
\end{align}
Here, $\gamma'$ is a parameter associated with the tunneling strength in the tunnel junction. 
While the non-local resistance increases linearly with increasing bias voltage $eV_S$, the noise increases instead quadratically. For finite temperature, we first numerically determine the $eV_S$-dependence of the magnon chemical potential $\mu_m$, and thereby we obtain the $eV_S$-dependence of the non-local resistance and noise. This finite temperature result is used to fit the experimental data for regime (ii) in Fig.~3(b).

In the limit of a long line junction, the last term in Eq.~\eqref{method:H} is modified to describe tunneling in the spatial region $0 \le x \le L$. In the equilibrated regime,  $L > \ell_{\text{eq}}$, this model yields non-local resistance and noise characteristics distinct from those in the single tunnel-junction model. 
Specifically, the equilibrated line-junction model predicts the following tunneling current, non-local resistance, and the excess noise in each individual absorption region,
\begin{align}
   I_{\text{ab}} &= \frac{e}{2h M} (|eV_S|- E_Z) \theta(|eV_S| - E_Z)\,, \label{supeq:linejunctioncurrent} \\ 
    \left | \frac{d V_{\text{ab}}}{dI} \right | &= \frac{h}{2 M e^2} \theta (|eV_S| - E_Z)\,, \label{supeq:linejunctionnlcond} 
\\ S_{\text{ab}} &= \frac{e^2}{h} ( T - T_0 )\,, \label{method:linejunctionnlnoise}
\end{align}
with the increased temperature $T$ of the system, Eq.~\eqref{eq:T}. Here $M = 5$ is the number of absorption regions. Notably, the non-local resistance \eqref{supeq:linejunctionnlcond} is constant in $eV_S$, whereas the noise [Eqs.~\eqref{eq:noise}-\eqref{eq:T}] instead increases linearly in $eV_S$ at sufficiently large bias voltage $eV_S$. In the calculation of Eqs.~\eqref{eq:T}, \eqref{supeq:linejunctioncurrent}, and \eqref{supeq:linejunctionnlcond}, we have assumed for simplicity that the magnons are absorbed in each individual absorption region with equal probabilities.
Note that the measured excess noise $S_I$ in Eq.~\eqref{eq:noise} has the additional factor $1/2$ compared with the excess noise generated in an absorption region, i.e., $S_I = \frac{1}{2} S_{\text{ab}}= 2 \times \frac{1}{4} S_{\text{ab}}$. The factor of $2$ reflects contributions from two noise spots (\textbf{`B'} and \textbf{`D'}) and the factor $1/4 = (1/2)^2$ originates from that only one channel out of the two emanating from the FC is measured at the bottom transverse contact, see Fig.~1(a).

We also calculate the dependence of the equilibration length $\ell_{\text{eq}}$ on the bias voltage $eV_S$. We do this by using the results for the partial-equilibration regime (short $L$) and inspecting at what $L$ the equilibration becomes strong. The result reads 
\begin{equation} \label{methods:equilibrationlength}
\ell_{\text{eq}} = \frac{1}{\gamma (|eV_S|-E_Z)},\,\quad \text{for }|eV_S| > E_Z\,.
\end{equation}
This equation implies a partial-equilibration regime for $|eV_S|$ slightly exceeding $E_Z$ and a strong-equilibration regime for larger $|eV_S|$, as discussed in the ``Discussion'' section above, and also illustrated in Fig.~3(b). Equation~\eqref{methods:equilibrationlength} shows that $\ell_{\text{eq}}$ increases significantly as the bias energy approaches $E_Z$, indicating that the equilibration process takes place very slowly near $|e V_S| \sim E_Z$. This happens because the absorption rate per unit length is proportional to $(|eV_S| -E_Z)^2$ [Eq.~\eqref{method:singlejunctioncurrent}] whereas the total tunneling current in the equilibrated regime scales as $(|eV_S| -E_Z)$ [Eq.~\eqref{supeq:linejunctioncurrent}].

\noindent\textbf{References}
\bibliography{Manuscript}

\section*{Acknowledgements}
A.D. thanks the Department of Science and Technology (DST) and Science and Engineering Research Board (SERB), India, for financial support (SP/SERB-22-0387) and acknowledges the Swarnajayanti Fellowship of the DST/SJF/PSA-03/2018-19. A.D. also thanks financial support from CEFIPRA: SP/IFCP-22-0005. S.K.S. and R.K. acknowledge the Prime Minister’s Research Fellowship (PMRF), Ministry of Education (MOE), and Inspire fellowship, DST for financial support, respectively. A.D.M., J.P., and Y.G. acknowledge support by the DFG Grant MI 658/10-2 and by the German-Israeli Foundation Grant I-1505-303.10/2019. Y.G. acknowledges support by the Helmholtz International Fellow Award, by the DFG Grant RO 2247/11-1, by CRC 183 (project C01), and by the Minerva Foundation. C.S. acknowledges funding from the 2D TECH VINNOVA competence Center (Ref. 2019-00068). This project has received funding from the European Union’s Horizon 2020 research and innovation programme under grant agreement No 101031655 (TEAPOT). K.W. and T.T. acknowledge support from the Elemental Strategy Initiative conducted by the MEXT, Japan and the CREST (JPMJCR15F3), JST.

\section*{Author contributions}
R.K., S.K.S., and U.R. contributed to device fabrication, data acquisition, and analysis. A.D. contributed to conceiving the idea and designing the experiment, data interpretation, and analysis. K.W. and T.T. synthesized the hBN single crystals. J.P., C.S., Y.G., and A.D.M. contributed to the development of theory and data interpretation, and all the authors contributed to writing the manuscript.

\section*{Competing interests}
The authors declare no competing interests.

\section*{Data and materials availability:}
The data presented in the manuscript are available from the corresponding author upon request. 

\newpage


\section*{Supplementary material}

# Supplementary information: Electrical noise spectroscopy of magnons in a quantum Hall ferromagnet

Ravi Kumar[1],[*] Saurabh Kumar Srivastav[1],[†] Ujjal Roy[1],[‡] Jinhong Park[2,3],[§] Christian Spånslätt[4], K. Watanabe[5], T. Taniguchi[5], Yuval Gefen[6], Alexander D. Mirlin[2,3], and Anindya Das[1],[¶]

[1]*Department of Physics, Indian Institute of Science, Bangalore, 560012, India.*

[2]*Institute for Quantum Materials and Technologies, Karlsruhe Institute of Technology, 76021 Karlsruhe, Germany.*

[3]*Institut für Theorie der Kondensierten Materie, Karlsruhe Institute of Technology, 76128 Karlsruhe, Germany.*

[4]*Department of Microtechnology and Nanoscience (MC2), Chalmers University of Technology, S-412 96 Göteborg, Sweden.*

[5]*National Institute of Material Science, 1-1 Namiki, Tsukuba 305-0044, Japan.*

[6]*Department of Condensed Matter Physics, Weizmann Institute of Science, Rehovot 76100, Israel.*

---

[*]equally contributed
[†]equally contributed
[‡]equally contributed
[§]equally contributed
[¶]anindya@iisc.ac.in

**This supplementary information contains the following details:**

**1. Device fabrication, characterization, and noise measurement setup**

**2. Noise and non-local resistance at $\nu = 1$ for various magnetic fields**

**3. Threshold voltage of magnon detection from noise and non-local resistance measurements**

**4. Temperature dependence of the non-local resistance**

**5. Contribution of phonons**

**6. Response of the bilayer graphene device**

**7. Noise of the bilayer graphene device**

**8. Gain and electron temperature estimation**

**9. Theoretical model**

- **9. 1. Key assumptions and hierarchy of energy scales**
- **9. 2. Model, electrical current, and noise**
- **9. 3. Current and noise in a single tunnel junction model**
- **9. 4. Current and noise in a line junction model**
- **9. 5. Noise generated in a line junction**
- **9. 6. Voltage dependence of equilibration length, overall behavior of the noise, and comparison to experimental data**

**Section S1: Device fabrication, characterization, and noise measurement setup**

Utilizing the dry transfer pick-up approach[1,2], we fabricated an encapsulated device consisting of a heterostructure involving hBN (hexagonal boron nitride), single-layer graphene (SLG), and graphite layers. The procedure for creating this heterostructure comprised mechanical exfoliation of hBN and graphite crystals onto an oxidized silicon wafer through the widely employed scotch tape method. Initially, a layer of hBN, with a thickness of approximately 25-30 nm, was picked up at a temperature of $90^oC$. This was achieved using a Poly-Bisphenol-A-Carbonate (PC) coated Polydimethylsiloxane (PDMS) stamp positioned on a glass slide, attached to a home-built micromanipulator. The hBN flake was then aligned over the previously exfoliated SLG layer, which was similarly picked up at $90^oC$. The subsequent step involved picking up the bottom hBN layer of similar thickness. Following the same process as above, this bottom hBN was picked up utilizing the previously acquired hBN/SLG assembly. Subsequent to this, the hBN/SLG/hBN heterostructure was employed to pick up the graphite flake. Ultimately, this resulting heterostructure (hBN/SLG/hBN/graphite) was placed on top of a 285 nm thick oxidized silicon wafer at a temperature of $180^oC$. To remove the residues of PC, this final stack was cleaned in chloroform (CHCl3) overnight, followed by cleaning in acetone and isopropyl alcohol (IPA). After this, Poly-methyl-methacrylate (PMMA) photoresist was coated on this heterostructure to define the contact regions using electron beam lithography (EBL). Apart from the conventional contacts, we defined a region of $\sim 6\mu m^2$ area in the middle of SLG flake, which acts as a floating metallic reservoir upon edge contact metallization. After EBL, reactive ion etching (with a mixture of CHF3 and O2 gas with a flow rate of 40 sccm and 4 sccm, respectively, at $25^oC$ with RF power of 60W) was used to define the edge contact[3]. The etching time was optimized such that the bottom hBN did not etch completely in order to isolate the contacts from the bottom graphite flake, which was used as the back gate. Finally, thermal deposition of Cr/Pd/Au (3/12/60 nm) was done in an evaporator chamber having a base pressure of $\sim 1 \times 10^{-7}$ mbar. After deposition, a lift-off procedure was performed in hot acetone and IPA. The device's schematics and measurement setup are shown in Fig. 1(a) in the main text. The distance from the floating contact to the ground contacts is $\sim 5\mu m$, whereas the transverse contacts are placed at a distance of $\sim 2.5\mu m$.

All measurements were done in a cryo-free dilution refrigerator with a $\sim$ 20mK base temperature. The electrical conductance was measured using the standard lock-in technique, whereas the noise was measured using an LCR resonant circuit at resonance frequency $\sim$ 740kHz. A schematic of the noise measurement setup is shown in Fig. 1(a) in the main text. The device was mounted on a chip carrier, which was connected to the homemade cold finger fixed to the mixing chamber plate of the dilution refrigerator. The ground contact (CG) pins were directly shorted to the cold finger in order to achieve the cold ground. The sample was current- biased during the measurements. Noise signals were amplified with a homemade cryogenic voltage pre-amplifier, which was thermalized to the 4K plate of the dilution refrigerator. This pre-amplified signal was then amplified using a voltage amplifier placed at the top of the fridge at room temperature. After the second stage of amplification, the amplified signal was measured using a spectrum analyzer (N9010A). All

noise measurements were done using a bandwidth ∼ 30 kHz. The resonant L//C tank circuit was built using an inductor $L$ of ∼ 365 µH made from a superconducting coil thermally anchored to the mixing chamber plate of the dilution refrigerator. A parallel capacitance $C$ of ∼ 125 pF develops along the coaxial lines connecting the sample to the cryogenic pre-amplifier.

We used two devices: one made of single layer graphene and one made of bilayer graphene. The quantum Hall response of the single-layer graphene is shown in Supplementary Fig. 1(a). It can be seen that robust QH plateaus are observed at the magnetic field strength $B = 1$T. Supplementary Figs. 1(b) and (c) show the temperature dependence of the longitudinal resistance $R_{XX}$ at $B = 1$T and 2T, respectively. The activation plots are shown in the insets, where we extract activation gaps of the $\nu = 1$ QH state. We estimate these gaps to ∼ 4K and 9K at $B = 1$T and $B = 2$T, respectively.

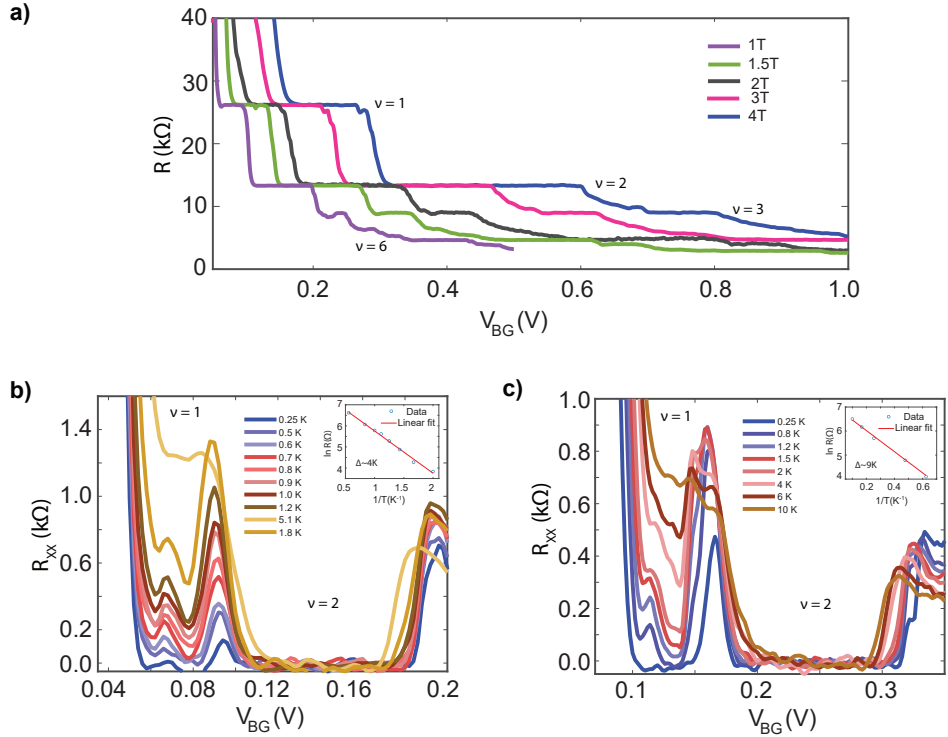

**Supplementary Figure 1: QH response of the single layer graphene device and activation gap.** (a) Hall response of the single-layer device at various magnetic field strengths. Panels (b) and (c) shows the $R_{XX}$ as a function of the gate voltage $V_{\text{BG}}$ at various temperatures for $B = 1$T and $B = 2$T, respectively. The insets show the activation gaps for $\nu = 1$.

4

**Section S2: Noise and non-local resistance at filling $\nu = 1$ for various magnetic fields**

In this section, we present the non-local resistance and noise measured at several magnetic field strengths across the $\nu = 1$ QH plateaus of the single-layer graphene device. The measurement schemes are described in the main text and shown in Fig. 1(a) in the main manuscript. Supplementary Fig. 2 summarizes the non-local resistance and noise data, from which the threshold voltages ($V_{\text{th}}$) are extracted and shown in Fig. 2(f) in the main manuscript. How the threshold voltage was extracted is presented in the next section.

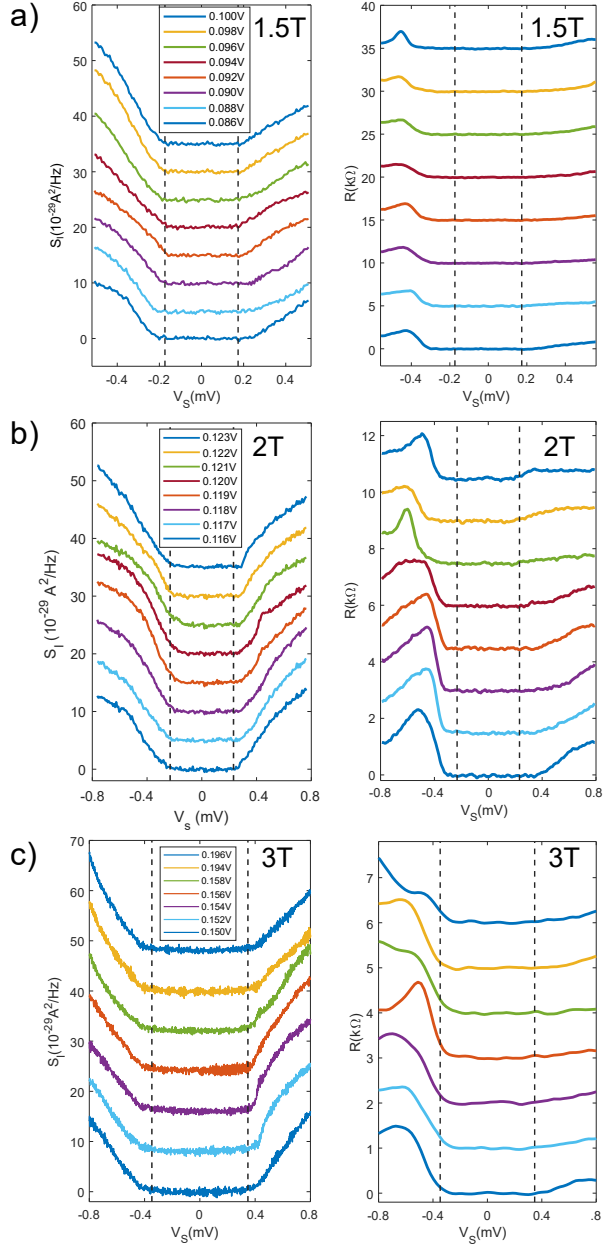

**Supplementary Figure 2: Noise and non-local resistance.** As described in the manuscript and shown in Fig. 1(a), the noise and the non-local resistance were measured. Panels (**a**), (**b**) and (**c**) show the measured noise (left column) and non-local resistance (right column) across the plateau of $\nu = 1$ at different magnetic fields ($B = 1.5$T, 2T and 3T) for single layer graphene device. The vertical dashed lines correspond to the Zeeman energy.

**Section S3: Threshold voltage of magnon detection from noise and non-local resistance measurements**

In this section, we describe how the threshold voltages ($V_{th}$) from the non-local resistance and noise data presented in Supplementary Fig. 2 were determined. Supplementary Fig. 3 shows one of the examples at different magnetic fields for both the non-local resistance and noise data. For the threshold voltage, we calculate the *rms* value of the data, and a sudden change in its magnitude is marked as a threshold voltage, which is shown by the vertical black lines in Supplementary Fig. 3. At a given magnetic field, the $V_{th}$ was extracted for several points across the plateau for the data set shown in Supplementary Fig. 2, and its mean value and standard deviation are shown in Fig. 2(f) in the main manuscript. For the noise data sets, we extract the $V_{th}$ for both the positive and negative bias voltages and take its mean. However, for non-local resistance data sets, we extract the $V_{th}$ for only the negative bias voltage side as the threshold voltage for the positive bias voltage side was not prominent.

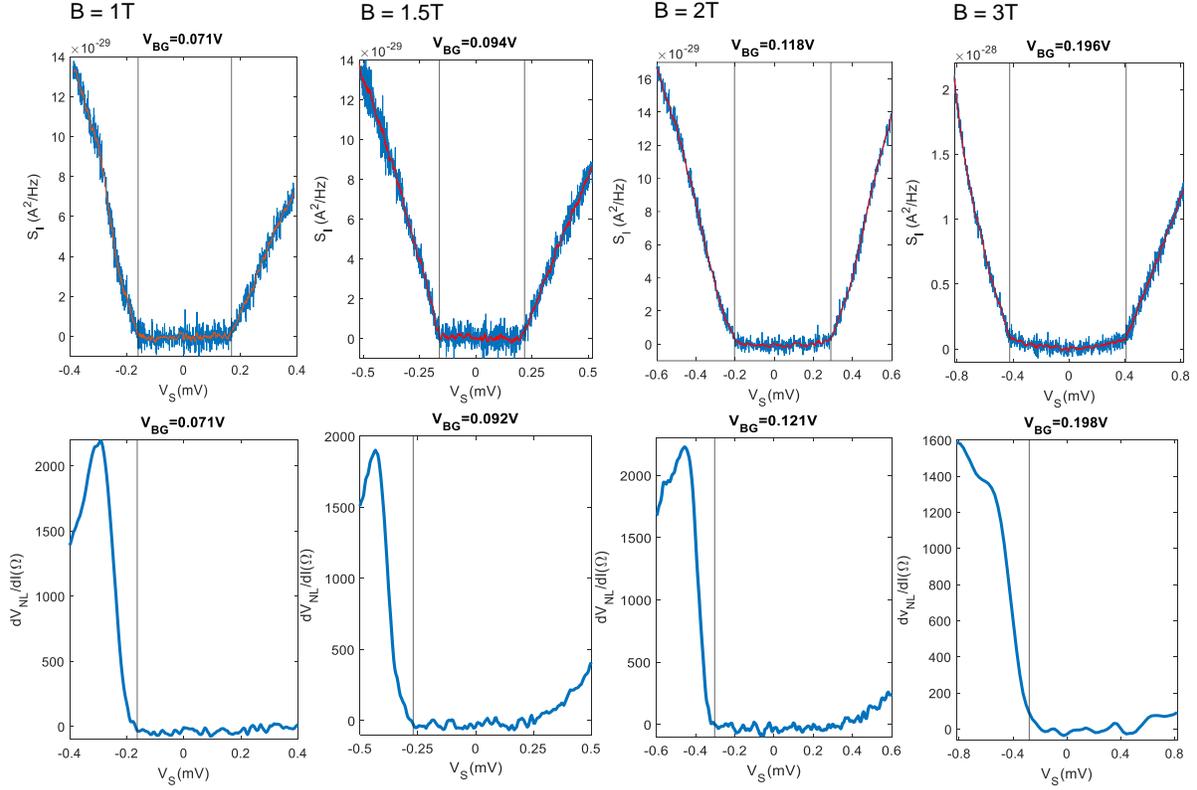

**Supplementary Figure 3: Threshold voltage.** To determine the threshold voltage from the noise data, we have taken a 10-point average of the raw data, shown by the red solid lines. We calculate the *rms* value of the average data, and a sudden change in its magnitude is marked as a threshold voltage, which is shown by the vertical black lines. We extract the threshold voltage for both the sign of bias voltages for the noise data and take its mean value. This exercise was done at several points of the plateau (points are shown in Supplementary Fig. 2), and its mean value and standard deviation for a given magnetic field are plotted in Fig. 2(f) in the main manuscript. A similar procedure was done for the non-local resistance data but only for the negative bias voltage side, as the transition for the positive bias voltage was hardly seen in most of the data.

### Section S4: Temperature dependence of the non-local resistance

In this section, we present the temperature dependence of the non-local resistance measured at the $\nu = 1$ QH plateau for $B = 1$T in the single-layer graphene device. As can be seen in Supplementary Fig. 4, the threshold voltage remains higher than the Zeeman energy until 200mK, and with increasing temperature the data gets broadened, and $V_{\text{th}}$ starts appearing at the vicinity of the Zeeman energy and even slightly lower

8

than the Zeeman energy at sufficiently high temperature. A similar temperature-broadening effect was observed in the threshold voltage for the non-local resistance at $B = 3$T but at the base temperature ($T_0 = 20$mK), as presented in Fig. 2(f) in the main manuscript. At a higher magnetic field ($B = 3$T), phonons play an important role since a larger current is required in order to generate magnons due to the higher Zeeman energy. Thus, the increased dissipation near '**A**' and '**E**' in Fig. 1(a) in the main manuscript is able to excite phonons even at the base temperature ($T_0 = 20$mK). The contribution of phonons is discussed further in Sec. S5. Note that no non-local resistance was detected even at elevated temperature ($T_0 = 600$mK) on the $\nu = 2$ QH plateau, which is expected due to its non-magnetic nature.

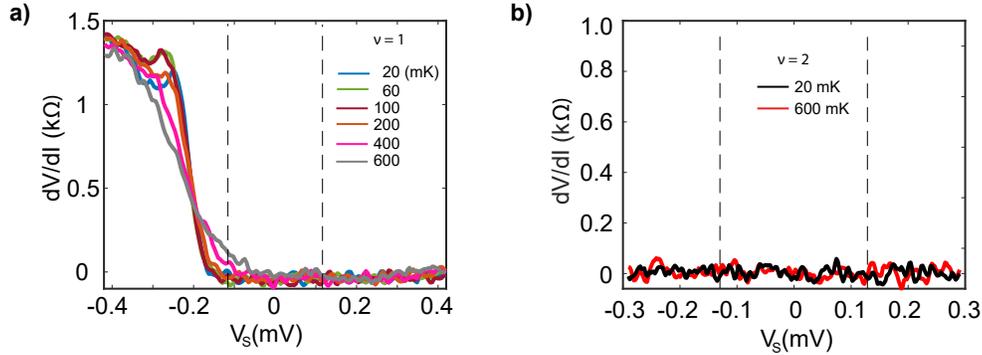

**Supplementary Figure 4: Temperature dependence of the non-local resistance** for (**a**) bulk filling $\nu = 1$, and (**b**) $\nu = 2$ at $B = 1$T. The non-local signal at $\nu = 1$ is almost the same up to 200 mK, and above 200 mK the data gets smoothed. No non-local signal was detected for $\nu = 2$ even at 600 mK as it is non-magnetic.

**Section S5: Contribution of phonons**

In this section, we present the contribution of phonons in our noise measurement for the single-layer graphene device. In order to know the contribution of phonons quantitatively, we measure the noise at the $\nu = 2$ non-magnetic QH plateau at several magnetic fields and at base temperature ($T_0 = 20$ mK). This is shown in Supplementary Fig. 5, where the x-axis is normalized with the Zeeman energy. Note that as described in Fig. 1(a) in the main manuscript, the application of a bias voltage at the source contact creates hot spots near '**A**' and '**E**', which can excite phonons and can travel through the bulk of the device and increase the temperature of the floating contact situated along the upstream direction, see Fig. 1(a) in the main manuscript. It can be seen from Supplementary Fig. 5 that at $B = 1$T and $B = 1.5$T, hardly any noise was detected. At $B = 2$T, a detectable noise from the phonon was measured, but its magnitude remains almost 6-7 times smaller than the noise measured for the $\nu = 1$ quantum Hall ferromagnetic phase, as shown in Fig. 2(e) in the main manuscript and Supplementary Fig. 2(b). At $B = 3$T, significant noise was detected from phonons, but its magnitude remains smaller than the noise measured for the $\nu = 1$ quantum

9

Hall ferromagnetic phase, as shown in Supplementary Fig. 2(c).

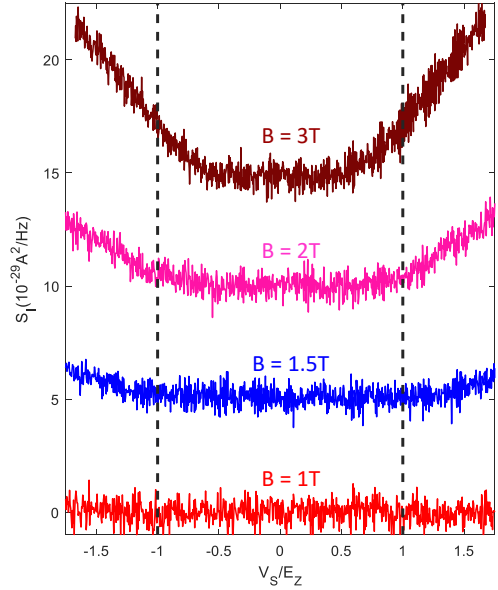

**Supplementary Figure 5: Noise measured at $\nu = 2$.** As described in the manuscript and shown in Fig. 1(a) in the main manuscript, the noise was measured along the upstream direction. Here, we show the measured noise at bulk filling $\nu = 2$ (which is non-magnetic) at different magnetic fields. For clarity, the data are vertically shifted by $5 \times 10^{-29}$ $A^2/Hz$. It can be seen that above $B = 3$T, significant noise due to phonons was detected. The bias voltages on the x-axes are normalized to the Zeeman energy, $E_Z$. The vertical dashed lines correspond to Zeeman energy. As mentioned in the main manuscript, at a higher magnetic fields, a larger current is required to reach $E_Z$, and thus stronger hot spots are created (points **'A'** and **'E'** in Fig. 1(a) in the manuscript), which excite phonons. However, most of the data was analyzed at $B = 1$T, where no phonon contribution was detected.

**Section S6: Response of the bilayer graphene device**

So far, we have presented and discussed the data obtained from the single-layer graphene device. In this section, we present the non-local resistance and noise measurements for our second device, made with bilayer graphene. For the bilayer graphene device, local doping is independently controlled by $SiO_2$ gating [shown by the gray filled region in Supplementary Fig. 6(a)], while the bulk filling of the device is controlled by the back graphite gate ["BG", shown as the sky blue filled region in Supplementary Fig. 6(a)]. Robust QH plateaus were observed at slightly higher magnetic fields, $B = 4$T, in contrast to $B = 1$T for the single-layer graphene device. This is shown in Supplementary Fig. 6(b), which depicts the resistance $R$ measured at the source contact as a function of the graphite back gate at $B = 4$T for the two cases where the $SiO_2$

gate was tuned at $\nu = 2$ and $\nu = 4$, respectively. Though the quality of the bilayer graphene device was not as good as the single-layer graphene device, the measured noise remained qualitatively similar to the single-layer device, as presented in the next section.

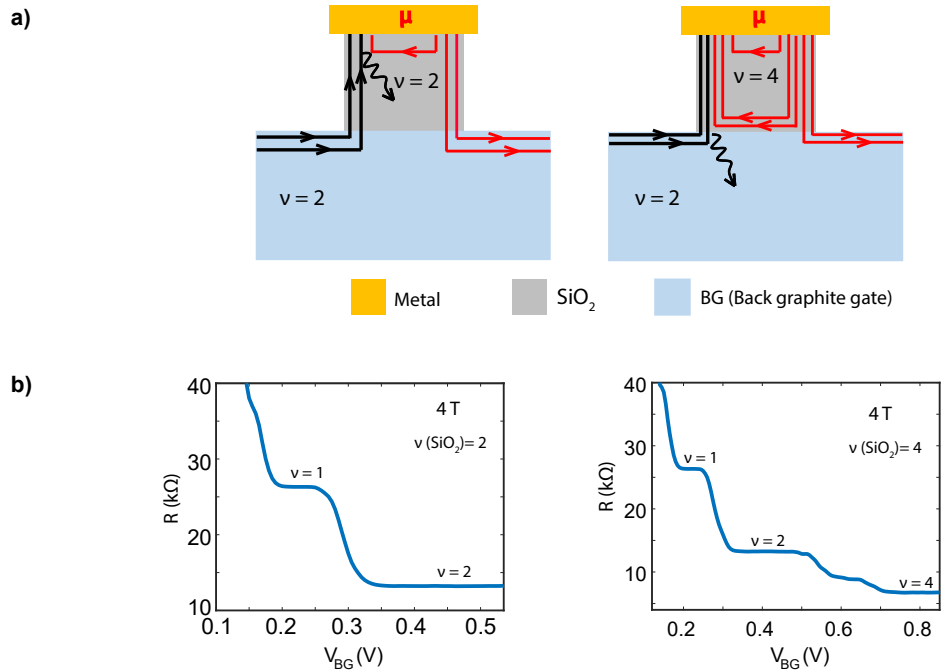

**Supplementary Figure 6: Controlled local doping near the ohmic contact and response of the bilayer graphene device.** For the bilayer graphene device, local doping is controlled by the $SiO_2$ (shown in gray) gating, while the device bulk filling is controlled by the back graphite gate (BG, shown in sky blue). (**a**) Left schematic: the bulk and local fillings were kept at $\nu = 2$. Right schematic: bulk and local fillings were kept at $\nu = 2$ and $\nu = 4$, respectively. The wavy lines indicate the magnon generation region for two different filling configurations. The corresponding QH responses as a function of the graphite back gate of the device are shown in panel (**b**) at $B = 4$T. For this measurement, the voltage was measured at the same source (current injected) contact.

### Section S7: Noise of the bilayer graphene device

In this section, we will discuss the noise measured for the bilayer graphene device. For bilayer graphene, $\nu = 2$ is a quantum Hall ferromagnet whereas $\nu = 4$ is non-magnetic[4,5] as shown schematically in Supplementary Fig. 7(b). The scheme for the noise measurement is shown in Supplementary Fig. 7(a) and discussed in details in the figure caption. Supplementary Fig. 7(c) shows a 2D color plot of the measured noise at $B = 4$T of the device, where the bulk filling is set to $\nu = 2$ while the filling near the contact is

11

kept at $\nu = 4$ (see Supplementary Fig. 6(a), right panel). The vertical dashed lines in Supplementary Fig. 7(c) correspond to the Zeeman energy and the solid magenta line corresponds to one of the line cuts. It can be seen that noise increases almost linearly for bias voltages above the Zeeman energy. However, it should be noted that there is finite noise measured even inside the Zeeman energy. This is not surprising as the measurement for the bilayer graphene device was done at $B = 4$T, and the phonon contribution was not negligible: As shown in Supplementary Fig. 7(d), there is finite noise even for the non-magnetic $\nu = 4$ QH state.

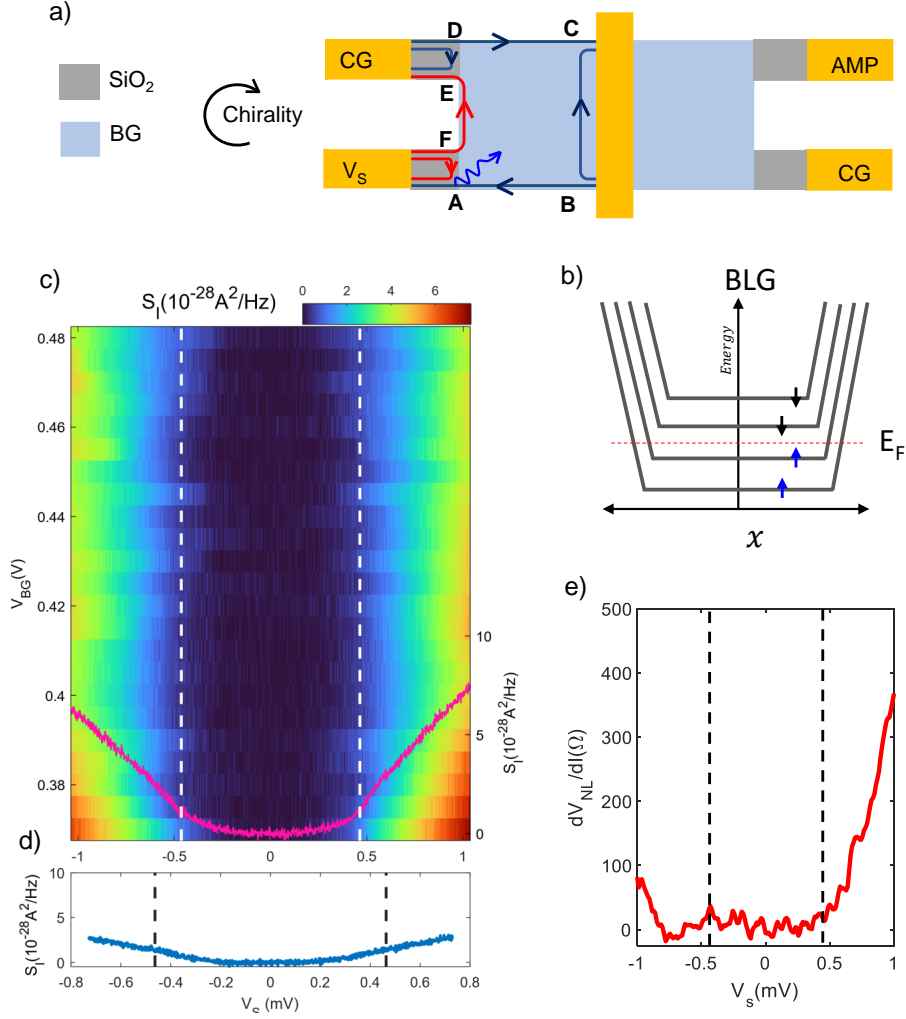

**Supplementary Figure 7: Measured noise** at $\nu = 2$. For bilayer graphene, $\nu = 2$ is a quantum Hall ferromagnet whereas $\nu = 4$ is non-magnetic[4,5] as shown in (**b**). (**a**) Device schematic for noise measurement. The $V_S$, CG, and AMP represent the voltage source, cold ground, and amplifier, respectively. The magnons are generated at points **'A'** and **'E'** for negative and positive bias voltages, respectively. In the schematic, it is shown only for negative bias voltage (by the wiggly line with an arrow). Note that in the schematic, each line with arrow corresponds to two edge modes for bilayer graphene (for bulk filling $\nu = 2$); for simplicity, we have shown a single line. The absorption at points **'B'** and **'D'** contribute to the noise and non-local resistance. (**c**) The 2D color plot of the measured noise (as described in Fig. 1(a) of the manuscript) at $B = 4$T of the device, where the bulk filling is set to $\nu = 2$ while the filling near the contact is kept at $\nu = 4$. The vertical dashed lines correspond to the Zeeman energy. The solid magenta line corresponds to one of the cut lines. It can be seen that noise increases almost linearly above the Zeeman energy. (**d**) Noise as a function of the bias voltage when both the filling of the bulk and contact are kept at $\nu = 4$. As expected, there is some contribution of noise from phonons. (**e**) Non-local resistance as a function of bias voltage. It can be seen that for negative bias voltage, the non-local signal is weaker as the distance from magnon generation point **'A'** to the absorption points **'B'** and **'D'** are almost equal.

13

## Section S8: Gain and electron temperature estimation

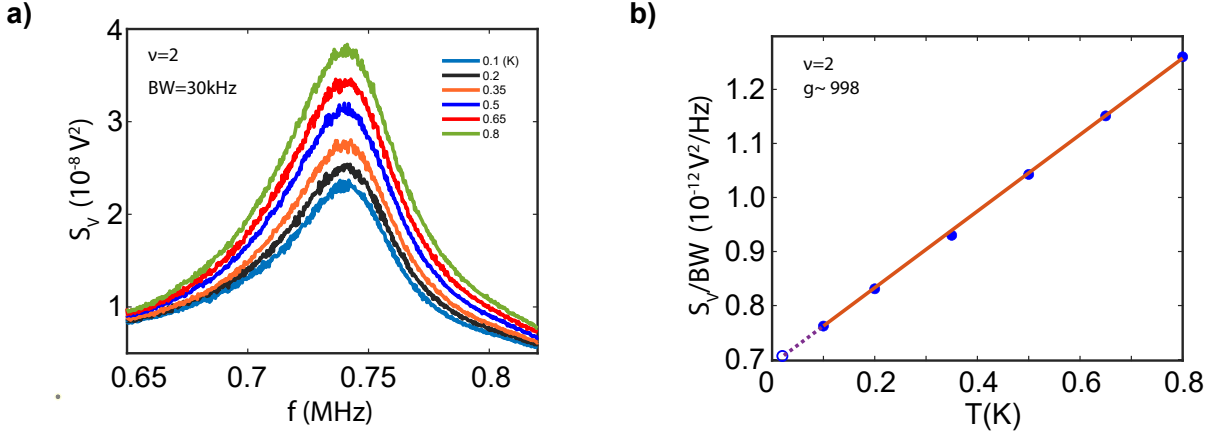

**Supplementary Figure 8: Gain and electron temperature estimation at $\nu = 2$.** (**a**) The voltage noise $S_V$ [see Eq. (S1)] measured by a spectrum analyzer is plotted as a function of the frequency at different bath temperatures for $\nu = 2$. From this plot, the resonance frequency of the tank circuit was found to be ∼740 kHz. (**b**) Blue solid circles represent the noise $S_V$ divided by bandwidth ($BW$) at the resonance frequency as a function of the bath temperature. The solid red line is a linear fit to the data from $0.1$ K to $0.8$ K and the dashed line is the linear extrapolation below $0.1$ K. Using Eq. (S2) and the slope of this linear fit, the gain $g$ was found to be equal to ∼998.

In our analysis of the noise data, it is crucial to know the gain of the amplifier chain and the electron temperature ($T_e$) of the system. In this section, we describe how we measure these two quantities. The electron temperature does not need to be the same as the bath temperature ($T_{bath}$) of the system (temperature of the MC plate in our experiment). We calculate precisely the gain and $T_e$ by measuring thermal noise at zero current bias.

We have estimated the gain of the amplification chain and the electron temperature from temperature-dependent Johnson-Nyquist noise (thermal noise). At zero impinging current, the equilibrium voltage noise spectrum is given by

$$S_V = g^2(4k_B T R + V_n^2 + i_n^2 R^2) BW, \tag{S1}$$

where $g$ is the total gain of amplification chain, $k_B$ the Boltzmann constant, $T$ the bath temperature (temperature of mixing chamber (MC) plate), $R$ is the resistance of quantum hall state, $V_n^2$ and $i_n^2$ are the intrinsic voltage and current noise of the amplifier, and $BW$ is the frequency bandwidth. The first term, $4k_B T R$ corresponds to the thermal noise. At a quantum Hall plateau, any change in bath temperature will only affect the first term in Eq. (S1), while all other terms are independent of temperature. If one plots the $\frac{S_V}{BW}$ as a function of temperature, the slope of the linear curve will be equal to $4g^2 k_B R$. Since at the quantum Hall

14

plateau, the resistance R is exactly known, one can easily extract the gain of the amplification chain from the slope and the intrinsic noise of the amplifier from the intercept. The gain is found using the following equation:

$$g = \sqrt{\left(\frac{\partial\left(\frac{S_V}{BW}\right)}{\partial T}\right)\left(\frac{1}{4k_B R}\right)}, \quad (S2)$$

where $\left(\frac{\partial\left(\frac{S_V}{BW}\right)}{\partial T}\right)$ is the slope of the linear fit. The implementation of this procedure is shown in Supplementary Fig. 8 for $\nu = 2$.

The noise spectrum ($S_V$) at zero impinging current measured on the $\nu = 2$ plateau at different bath temperatures is shown as a function of frequency in Supplementary Fig. 8(a). The $S_V$ value at the resonance frequency, divided by BW, is plotted as a function of bath temperature in Supplementary Fig. 8(b), where the red solid line is the linear fit to the data in the temperature range from 0.1K to 0.8K. From the slope, we extract the gain, which is found to be $\sim 998$. Note that we do not use the base temperature data for the fitting in Supplementary Fig. 8(b), because the electron temperature ($T_e$) could be different from the base temperature.

As the gain is known, one can calculate the $(V_n^2 + i_n^2 R^2)$ from the intercept of the linear fitting of $S_v/BW$ vs temperature. Now from the known value of the measured noise at the base temperature, the corresponding electron temperature ($T_e$) can be found directly using the following equation:

$$T_e = \frac{\left(\left(\frac{S_V}{g^2 BW}\right) - (V_n^2 + i_n^2 R^2)\right)}{4k_B R}. \quad (S3)$$

The measured value of noise for $\nu = 2$ at base temperature corresponds to $T_e = 23$ mK, which is consistent with the electron temperature measured in our previous work[6,7]. The fact that $T_e$ is very close to the bath temperature can be also seen directly from Supplementary Fig. 8(b): The $T_{bath} = 20$ mK data point (blue open circle) is located almost on the dashed line representing the extrapolation of the linear fit into the region below 0.1 K.

**Section S9: Theoretical model**

In this section, we calculate the electrical current and noise generated in the magnon absorption regions. We perform this calculation with two different models describing the magnon generation and absorption regions ('**A**', '**B**', '**C**', '**D**', '**E**', and '**F**', see Fig. 1(a) in the main text): (i) a single tunnel junction model and (ii) a line junction model. In the single tunnel junction model, we assume that the equilibration between edge channels and magnons in those regions is only partial. In other words, the characteristic equilibration length scale $\ell_{\rm eq}$ is much larger than the physical length $L$ of the regions. In contrast, the line junction model assumes $\ell_{\rm eq} < L$, i.e., full equilibration. As we show below, the two models yield distinct electric current and noise characteristics. We further show that the first model is applicable in a range of bias voltages $|eV_S|$ sufficiently close to $E_Z$, while the second model is justified for larger $|eV_S|$. Finally, we compare these theoretical predictions with our experimental results.

A word of caution concerning the terminology is in order at this point. Strictly speaking, the length $\ell_{\rm eq}$ corresponds to a pre-equilibration. The true equilibration at low temperatures $T \ll E_Z$ takes place at an exponentially large length and is thus not relevant experimentally. We discuss this point in more detail below in subsection S9.4.

This theoretical section is organized as follows. In subsection S9. 1, we establish the key assumptions of the calculations and the hierarchy of involved energy scales in the system. In subsections S9. 2 and S9. 3, we calculate the electrical current and noise in the single tunnel junction model. The following subsection S9. 4 presents the line junction model and its associated current and noise characteristics. Technical details of the calculation of noise are presented in S9. 5. In the last subsection S9. 6, we calculate the equilibration length, determine ranges of validity of both models, summarize our theoretical predictions, and compare them with the experimental data.

**S9.1. Key assumptions and hierarchy of energy scales.** Before diving into the actual calculation, we first establish the hierarchy of involved energy scales in the system. The relevant energy scales are the bias energy $|eV_S|$, the Zeeman energy $E_Z$ (which is the minimal energy of a magnon), the total inter-edge channel electron tunneling rate $\Gamma$ in the absorption regions (or, equivalently, the total magnon absorption rate), the thermal energy $k_B T_0$, and the level spacing $\delta E$ of the magnonic spectrum. In our calculations, we assume the following hierarchy:

$$\delta E < k_B T_0 \ll E_Z < |eV_S|. \tag{S4}$$

From the condition $k_B T_0 \ll E_Z$, we can neglect any generation of magnons by thermal excitations, so that the magnon generation is almost entirely due to tunneling of electrons between edge channels. In the regime $|eV_S| < E_Z$, this tunneling process, and hence the magnon generation, is strongly suppressed since such a process does not satisfy energy conservation (up to exponentially small contributions). By contrast, for

biases $|eV_S| > E_Z$, the tunneling is efficient, resulting in significant magnon generation. Importantly, in this bias regime, *real* processes of creation of magnon states take place, such that the generated magnons remain in the bulk over a relatively long time scale. During their life-time, the magnons may experience inelastic scattering processes, which give rise to an equilibrium distribution of the bulk magnons. This distribution is characterized by an effective temperature of $T$, to be derived below. We assume that the inelastic magnon relaxation rate is much larger than the inverse dwell time of magnons in the quantum Hall bulk, i.e.,

$$\Gamma_{\text{in}} \gg \frac{\hbar}{t_{\text{dwell}}} \approx \min\left(\Gamma, \frac{\hbar v_m}{L_{\text{flight}}}\right). \tag{S5}$$

Here, $v_m$ is the magnon velocity, and the typical length scale $L_{\text{flight}}$ of the magnon propagation is the geometric size of the quantum Hall bulk. Under the assumption (S5), transitions between magnonic states are possible and thus the magnons lose coherence when they propagate before eventually decaying in one of the absorption regions. This loss of coherence permits us to treat the magnon generation or absorption regions independently of each other. This independence of tunneling processes in different regions is one of key assumptions of our theoretical analysis presented in detail below. Finally, the condition $\delta E < k_B T_0$ in Eq. (S4) is required in order to treat the magnonic spectrum as continuous.

**S9.2. Model, electrical current, and noise.** In this subsection, we consider a magnon generation and absorption process to take place in the vicinity of the contacts, **'A', 'B', 'C', 'D', 'E'**, and **'F'**, see Fig. 1(a) in the main text. By treating those regions as a single tunnel junction, we derive general formulas for the electrical current and noise generated in the junction. The main formulas will be used in subsection S9.3. Furthermore, the general formulas of the presented subsection can be straightforwardly extended to the case of a line junction (equilibrated regime), as is done in subsection S9.4, where the current and the noise in that regime are evaluated.

Each of the magnon generation and absorption regions consists of two co-propagating edge channels with opposite spins $s = \uparrow, \downarrow$. The Hamiltonian for these channels, the bulk magnons, and their coupling is $\hat{H} = \hat{H}_0 + \hat{H}_T + \hat{H}_m$, where $\hat{H}_0$ describes the kinetic energy for unbiased edge channels

$$\hat{H}_0 = -iv \sum_{s=\uparrow,\downarrow} \int dx \, \psi_s^\dagger(x) \partial_x \psi_s(x). \tag{S6}$$

Here, $\psi_s^\dagger(x)$ creates an electron at position $x$ with spin $s = \uparrow, \downarrow$ and both channels have velocity $v$. The bulk of the $\nu = 1$ quantum Hall states is spin-polarized with spin-$\uparrow$ electrons. The lowest-lying excitations in the bulk are spin excitations associated with spin flips ($\Delta S_z = \hbar$) called *magnons* [8]. The magnon excitation has a quadratic dispersion $\omega_{\bm{q}} \propto \bm{q}^2$ (with $\bm{q}$ the magnon momentum) and has an energy gap set by the Zeeman energy $E_Z$. The magnon Hamiltonian is thus

$$\hat{H}_m = \sum_{\bm{q}} (E_Z + \hbar \omega_{\bm{q}}) b_{\bm{q}}^\dagger b_{\bm{q}} \tag{S7}$$

17

in which $b_q^\dagger$ ($b_q$) creates (destroys) a magnon excitation with momentum $q$. Due to conservation of angular momentum, electron tunneling from the inner edge channel with spin-↓ to the outer one with spin-↑ requires generation of magnons. This effect is captured by the tunneling Hamiltonian

$$\hat{H}_T = W\psi_\uparrow^\dagger(x=0)\psi_\downarrow(x=0,t)b^\dagger(x=0) + \text{h.c.}. \tag{S8}$$

Here, we assumed that tunneling only occurs at a single tunnel junction (taken at position $x = 0$); a generalization of $H_T$ to the case of a line junction is straightforward. The amplitude for the tunneling is $W$. We next include the effect of the bias voltage $V_S$ by introducing a time dependence in $H_T(t)$ through a gauge transformation [9]. This transformation leads to

$$\hat{H}_T(t) = W\psi_\uparrow^\dagger(0,t)\psi_\downarrow(0,t)b^\dagger(0,t)e^{-\frac{i}{\hbar}eV_St} + \text{h.c.}. \tag{S9}$$

Importantly, we have $eV_S > 0$ and $eV_S = 0$ describing the magnon generation and absorption regions, respectively. In other words, magnon generation is induced by a bias voltage (and in fact requires a sufficiently large voltage to overcome the magnon gap, as discussed above and will be explicit in calculation below). On the other hand, the absorption regions are unbiased.

From the Heisenberg equation of motion, we derive the tunneling current operator, given by

$$\hat{I}(t) = -\frac{ie}{\hbar}\left(W\psi_\uparrow^\dagger(0,t)\psi_\downarrow(0,t)b^\dagger(0,t)e^{-\frac{i}{\hbar}eV_St} - \text{h.c.}\right). \tag{S10}$$

We obtain the expectation value of the current at time $t$ with the Keldysh technique: [9]

$$I(t) \equiv \langle\hat{I}(t)\rangle = \frac{1}{2}\sum_{\eta=\pm}\langle T_C\hat{I}(t^\eta)e^{-\frac{i}{\hbar}\int_C dt\hat{H}_T(t)}\rangle. \tag{S11}$$

Here, $\langle...\rangle$ denotes averaging over the ground state of $H_0 + H_m$ and $C$ denotes the Keldysh contour, consisting of a forward (backward) branch to evolve time from $-\infty(\infty)$ to $\infty(-\infty)$. The argument $t^\eta$ denotes time $t$ on the forward (backward) branch, parameterized by $\eta = \pm$. We assume that the tunneling is weak, $W \ll 1$, and thus keep terms only up to second order in the tunneling amplitude $W$. In this approximation, the expectation value of the current becomes

$$I(t=0) = -\frac{ie|W|^2}{2\hbar^2}\sum_{\eta,\eta_1=\pm 1}\eta_1\int_{-\infty}^{\infty}dt_1\Big(e^{\frac{i}{\hbar}eV_St_1}G_\uparrow^{\eta_1\eta}(t_1,0)G_\downarrow^{\eta\eta_1}(0,t_1)D^{\eta_1\eta}(t_1,0)$$
$$- e^{-\frac{i}{\hbar}eV_St_1}G_\uparrow^{\eta\eta_1}(0,t_1)G_\downarrow^{\eta_1\eta}(t_1,0)D^{\eta\eta_1}(0,t_1)\Big). \tag{S12}$$

Here, $G_s^{\eta_1\eta_2}(t_1,t_2) \equiv -i\langle T_C\psi_s(t_1^{\eta_1})\psi_s^\dagger(t_2^{\eta_2})\rangle$ is the fermionic Keldysh Green functions for the edge channel with spin $s = \uparrow,\downarrow$, while $D^{\eta_1\eta_2}(t_1,t_2) \equiv -i\langle T_Cb(t_1^{\eta_1})b^\dagger(t_2^{\eta_2})\rangle$ is the bosonic Keldysh Green functions for the magnons. By using the relation $G_s^{\eta_1\eta_2}(t_1,t_2) = -G_s^{\eta_2\eta_1}(t_2,t_1)$ (which follows from particle-hole

18

symmetry in the edge channel spectrum), Eq. (S12) reduces to

$$I(t=0) = -\frac{ie|W|^2}{2\hbar^2} \sum_{\eta,\eta_1=\pm 1} \eta_1 \int_{-\infty}^{\infty} dt_1 G_\uparrow^{\eta_1\eta}(t_1,0) G_\downarrow^{\eta\eta_1}(0,t_1)$$
$$\times \left(e^{\frac{i}{\hbar}eV_St_1} D^{\eta_1\eta}(t_1,0) - e^{-\frac{i}{\hbar}eV_St_1} D^{\eta\eta_1}(0,t_1)\right). \tag{S13}$$

Since $G_s^{\eta\eta}(t)$ is an even function of $t$, the contribution from $\eta_1 = \eta$ vanishes and Eq. (S13) is further simplified as

$$I = \frac{ie|W|^2}{\hbar^2} \int_{-\infty}^{\infty} dt_1 G_\uparrow^>(t_1) G_\downarrow^<(-t_1) \left(D^>(t_1) e^{\frac{i}{\hbar}eV_St_1} - D^<(-t_1) e^{-\frac{i}{\hbar}eV_St_1}\right)$$
$$= \frac{ie|W|^2}{\hbar^2} \int \frac{d\omega_1}{2\pi} \int \frac{d\omega_2}{2\pi} \left(G_\uparrow^>(\omega_1) G_\downarrow^<(\omega_2 - \frac{eV_S}{\hbar}) D^>(\omega_2 - \omega_1)\right.$$
$$\left. - G_\uparrow^<(\omega_1) G_\downarrow^>(\omega_2 - \frac{eV_S}{\hbar}) D^<(\omega_2 - \omega_1)\right). \tag{S14}$$

Equation (S14) is the central result of this subsection and will be used in subsection S9.3 for the tunnel-junction model and, with an appropriate modification, in subsection S9.4 for the line-junction model.

We now use the same formalism to obtain the zero frequency noise. The noise $S(t)$ is given by

$$S(t) \equiv \langle \hat{I}(t)\hat{I}(0)\rangle + \langle \hat{I}(0)\hat{I}(t)\rangle - 2\langle \hat{I}(t)\rangle^2$$
$$= \sum_{\eta=\pm} \langle T_C \hat{I}(t^\eta) \hat{I}(0^{-\eta}) e^{-\frac{i}{\hbar} \int_C dt \hat{H}_T(t)}\rangle - 2\langle \hat{I}(t)\rangle^2. \tag{S15}$$

To second order in the tunneling strength, the zero frequency noise reads

$$S = \int_{-\infty}^{\infty} dt S(t) \simeq \sum_{\eta=\pm} \int_{-\infty}^{\infty} dt \langle T_C \hat{I}(t^\eta) \hat{I}(0^{-\eta})\rangle$$
$$= \frac{2ie^2|W|^2}{\hbar^2} \int_{-\infty}^{\infty} dt e^{-\frac{i}{\hbar}eV_St} (G_\uparrow^>(-t) G_\downarrow^<(t) D^>(-t) + G_\uparrow^<(-t) G_\downarrow^>(t) D^<(-t))$$
$$= \frac{2ie|W|^2}{\hbar^2} \int \frac{d\omega_1}{2\pi} \int \frac{d\omega_2}{2\pi} \left(G_\uparrow^>(\omega_1) G_\downarrow^<(\omega_2 - \frac{eV_S}{\hbar}) D^>(\omega_2 - \omega_1)\right.$$
$$\left. + G_\uparrow^<(\omega_1) G_\downarrow^>(\omega_2 - \frac{eV_S}{\hbar}) D^<(\omega_2 - \omega_1)\right). \tag{S16}$$

By using the well-known expressions for the equilibrium Green functions of edge channels and magnons (both at temperature $T_0$), we obtain the following relation between the tunneling current $I$ and the zero frequency noise $S$:

$$S = 2eI \coth\left(\frac{eV_S - \mu_m}{2T_0}\right). \tag{S17}$$

Here, $\mu_m$ is the chemical potential of the magnons. Note that magnons have a non-zero chemical potential since we consider a non-equilibrium situation (finite voltage $eV_S$). Equation (S17) bears similarity to a usual formula for (zero-frequency) noise in tunnel junctions. The difference is in chemical potential $\mu_m$ entering Eq. (S17). This is because an electron tunneling in our problem involves creation or annihilation of a magnon.

**S9.3. Current and noise in a single tunnel junction model.** In this subsection, we consider a single tunnel junction model, which was described above. Within this model, we compute the electrical current and noise generated in the absorption regions by employing Eqs. (S14) and (S17). For generality, we allow for tunneling amplitude $W$ to be different in the generation and absorption regions, as indicated by subscripts "ge" and "ab" below.

As discussed above, we treat the tunneling processes in the magnon generation and absorption regions independently of each other and thus calculate the electrical current in each region separately. Applying Eq. (S14), the electrical current in the magnon generation region is given as

$$
\begin{aligned}
I_{\text{ge}} = \frac{2\pi}{\hbar} e |W_{\text{ge}}|^2 \rho_{\text{edge}}^2 \rho_m \int d\omega_1 \int d\omega_2 \theta(\omega_2 - \omega_1 - E_Z) \\
\times \big[ (1 - f(\omega_1)) f(\omega_2 - |eV_S|)(1 + b(\omega_2 - \omega_1 - \mu_m)) \\
- f(\omega_1)(1 - f(\omega_2 - |eV_S|)) b(\omega_2 - \omega_1 - \mu_m) \big].
\end{aligned} \quad (S18)
$$

Here, we have used the equilibrium Green's functions for the edge channels, $G^>_{\uparrow/\downarrow}(\omega) = -2\pi i \rho_{\text{edge}}(1 - f(\omega))$ and $G^<_{\uparrow/\downarrow}(\omega) = 2\pi i \rho_{\text{edge}} f(\omega)$, as well as the Green's function for the magnons

$$
\begin{aligned}
D^>(\omega) &= -2\pi i \sum_{i_N, f_{N+1}} \sum_{\boldsymbol{q}} W_{i_N} \langle i_N | b_{\boldsymbol{q}} | f_{N+1} \rangle \langle f_{N+1} | b_{\boldsymbol{q}}^\dagger | i_N \rangle \delta(\omega - (E_Z + \omega_{\boldsymbol{q}})) \\
&= -2\pi i \sum_{\boldsymbol{q}} (1 + b(\omega_{\boldsymbol{q}} + E_Z - \mu_m)) \delta(\omega - (E_Z + \omega_{\boldsymbol{q}})) \\
&= -2\pi i \rho_m \theta(\omega - E_Z)(1 + b(\omega - \mu_m)).
\end{aligned} \quad (S19)
$$

The density of states $\rho_{\text{edge}}$ and $\rho_m$ for the edge modes and magnons are assumed to be energy independent. In Eq. (S19), $|i_N\rangle$ ($|f_N\rangle$) is an initial (final) Fock state with $N$ magnons. In the second equality, we used that only states with $|f_{N+1}\rangle \propto b_{\boldsymbol{q}}^\dagger |i_N\rangle$ survive in the summation. Furthermore, $W_{i_N}$ is the probability that the initial state is $|i_N\rangle$, and $\langle i_N | b_{\boldsymbol{q}} | f_{N+1} \rangle \langle f_{N+1} | b_{\boldsymbol{q}}^\dagger | i_N \rangle$ is further expressed as $\langle i_N | b_{\boldsymbol{q}} | f_{N+1} \rangle \langle f_{N+1} | b_{\boldsymbol{q}}^\dagger | i_N \rangle = (1 + b(\omega_{\boldsymbol{q}} + E_Z))$. This equality is based on that the magnon states follow a Bose distribution function $b(\omega - \mu_m) = 1/(e^{\beta(\omega - \mu_m)} - 1)$ with the magnonic chemical potential $\mu_m$. The electrical current (S18) can now be written as

$$
I_{\text{ge}} = -e(\Gamma^{\text{ge}}_{N+1, N} - \Gamma^{\text{ge}}_{N-1, N}), \quad (S20)
$$

in terms of the rate $\Gamma^{\text{ge}}_{N+1,N}$ ($\Gamma^{\text{ge}}_{N-1,N}$) at which the magnon number increases (respectively, decreases) by one. Similarly, the current generated in the absorption regions can be written as

$$I_{\text{ab}} = -e(\Gamma^{\text{ab}}_{N+1,N} - \Gamma^{\text{ab}}_{N-1,N}). \tag{S21}$$

The rates in Eq. (S20) are given as

$$\begin{aligned}\Gamma^{\text{ge}}_{N+1,N} &= -\frac{2\pi}{\hbar}|W_{\text{ge}}|^2 \rho^2_{\text{edge}}\rho_m \int d\omega_1 \int d\omega_2 \theta(\omega_2 - \omega_1 - E_Z) \\ &\quad \times (1-f(\omega_1))f(\omega_2 - |eV_S|)(1 + b(\omega_2 - \omega_1 - \mu_m)) \\ &= -\frac{2\pi}{\hbar}|W_{\text{ge}}|^2 \rho^2_{\text{edge}}\rho_m \int d\omega(\omega - |eV_S|)(1 + b(\omega - \mu_m))b(\omega - |eV_S|)\theta(\omega - E_Z) \\ &= -\frac{\gamma_{\text{ge}}}{h} \int d\omega(\omega - |eV_S|)(1 + b(\omega - \mu_m))b(\omega - |eV_S|)\theta(\omega - E_Z),\end{aligned} \tag{S22}$$

and

$$\Gamma^{\text{ge}}_{N-1,N} = \frac{\gamma_{\text{ge}}}{h} \int d\omega(\omega - |eV_S|)b(\omega - \mu_m)b(|eV_S| - \omega)\theta(\omega - E_Z). \tag{S23}$$

Here, $\gamma_{\text{ge}}$ is defined as $\gamma_{\text{ge}} \equiv (2\pi)^2|W_{\text{ge}}|^2 \rho^2_{\text{edge}}\rho_m$ and is assumed to be constant in energy. Similarly, the rates in Eq. (S21) read

$$\begin{aligned}\Gamma^{\text{ab},i}_{N+1,N} &= -\frac{\gamma^{\text{ab},i}}{h} \int d\omega \omega(1 + b(\omega - \mu_m))b(\omega)\theta(\omega - E_Z), \\ \Gamma^{\text{ab},i}_{N-1,N} &= \frac{\gamma^{\text{ab},i}}{h} \int d\omega \omega b(\omega - \mu_m)b(-\omega)\theta(\omega - E_Z).\end{aligned} \tag{S24}$$

The superscript $i$ labels here the five absorption regions shown in Figs. 1(a)-(b) in the main text. Combining Eqs. (S20)-(S24), we obtain the current generated in the magnon creation and absorption processes as

$$I_{\text{ge}} = \frac{e\gamma^{\text{ge}}}{h} \int_{E_Z}^{\infty} d\omega(\omega - |eV_S|)(b(\omega - |eV_S|) - b(\omega - \mu_m)) \tag{S25}$$

$$I^i_{\text{ab}} = \frac{e\gamma^{\text{ab},i}}{h} \int_{E_Z}^{\infty} d\omega \omega(b(\omega) - b(\omega - \mu_m)) . \tag{S26}$$

Note that these rates are independent of the number of magnons.

In the steady state, the number of magnons does not change in time and hence

$$I_{\text{ge}} = -\sum_{i=1}^{M} I^i_{\text{ab}} \tag{S27}$$

should hold. Here, $M = 5$ is the number of the absorption regions (see Figs. 1(a)-(b) in the main text). By using Eq. (S27), we can self-consistently determine the effective magnon chemical potential $\mu_m$. We further assume that the magnon and the edge channels are all at the system base temperature $T_0$.

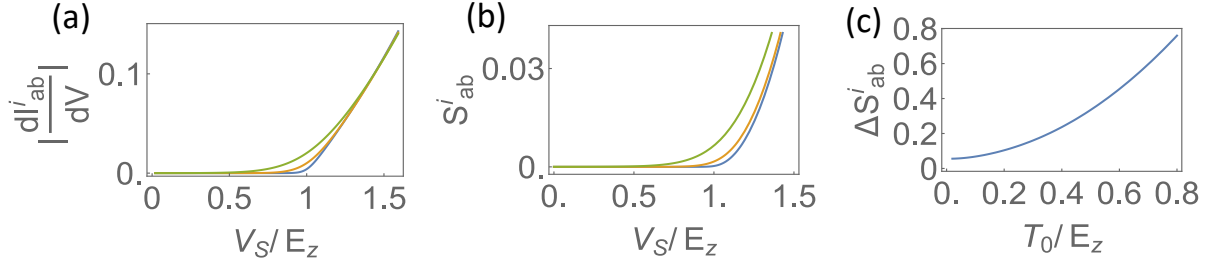

**Supplementary Figure 9: Singe tunnel junction results:** (a) Non-local differential conductance $\frac{dI^i_{ab}}{dV}$ and (b) noise $S^i_{ab}$ generated at an absorption region as a function of the bias voltage $V$ at different temperatures $T_0$. Blue lines: $T_0 = 0.02 E_Z$, orange lines: $T_0 = 0.05 E_Z$, green lines: $T_0 = 0.1 E_Z$, where $E_Z$ is the Zeeman energy (magnon energy gap). (c) Excess noise $\Delta S^i_{ab}$ vs temperature $T_0$ at fixed bias energy $eV_S = 1.5 E_Z$. With increasing temperature, the excess noise monotonically increases as more magnons are generated, since a larger phase space becomes available at higher temperatures. Both differential conductance and noise are plotted in units of $\frac{e^2}{h}\gamma^{ge}$. We set $e = k_B = 1$ for the plots.

The tunneling current generated in each individual absorption region is given by

$$I^i_{ab} = \frac{\gamma^{ab,i}}{\gamma^{ab}} \sum_{i=1}^{M} I^i_{ab} = -\frac{\gamma^{ab,i}}{\gamma^{ab}} I_{ge}, \tag{S28}$$

where $\gamma^{ab} \equiv \sum_{i=1}^{M} \gamma^{ab,i}$. At zero temperature, Eq. (S25) takes the simplified form

$$I_{ge} \approx \frac{e\gamma^{ge}}{2h}(|eV_S| - E_Z)^2 \theta(|eV_S| - E_Z), \tag{S29}$$

resulting in $I^i_{ab}$ and the corresponding non-local conductance from Eq. (S28) as

$$I^i_{ab} \approx -\frac{\gamma^{ab,i}}{\gamma^{ab}} \frac{e\gamma^{ge}}{2h}(|eV_S| - E_Z)^2 \theta(|eV_S| - E_Z), \tag{S30}$$

$$\frac{dI^i_{ab}}{dV} \approx -\frac{\gamma^{ab,i}}{\gamma^{ab}} \frac{e^2}{h}\gamma^{ge}(|eV_S| - E_Z)\theta(|eV_S| - E_Z). \tag{S31}$$

By using Eq. (S17), we obtain the noise generated at an absorption region

$$S^i_{ab} = 2e \coth\left(\frac{\mu_m}{2T_0}\right) |I^i_{ab}| \xrightarrow{\mu_m/T_0 \to \infty} \frac{e^2}{h} \frac{\gamma^{ab,i}}{\gamma^{ab}} \gamma^{ge}(|eV_S| - E_Z)^2 \theta(|eV_S| - E_Z). \tag{S32}$$

By self-consistently calculating $\mu_m$ from the steady state condition Eq. (S27), we numerically obtain $I^i_{ab}$, $\frac{dI^i_{ab}}{dV}$, and $S^i_{ab}$. We plot $\frac{dI^i_{ab}}{dV}$, and $S^i_{ab}$ as functions of the bias voltage at finite temperature in Supplementary Figs. 9(a)-(b). While the non-local conductance increases linearly in the bias voltage $|eV_S| - E_Z \gg T_0$, the noise increases quadratically. These characteristic behaviors of the non-local conductance and the noise are consistent with the zero temperature results, see Eqs. (S31) and (S32). The excess noise $\Delta S^i_{ab}$ vs $T_0$ at

22

fixed bias is plotted in Supplementary Fig. 9(c). It is seen that the excess noise increases with increasing temperature $T_0$ in the single tunnel-junction model. This happens because more magnons are generated due to the larger magnonic phase space available at higher temperatures. This behavior is at variance with the experimental dependence of noise on temperature for a voltage $|eV_S|$ well above $E_Z$, see Fig. 3(e) in the main text. As we show below, at such voltages a different model—that of line junction corresponding to full equilibration—is applicable, which explains the experimental behavior of noise in Fig. 3(e).

**S9.4. Current and noise in a line junction model.** In contrast to the previous subsection, we assume here that magnon-generation and absorption processes occur along a line junction rather than in a single tunnel junction. We model such a line junction as an extended segment of length $L$ with two co-propagating edge channels. Along this segment electrons tunnel incoherently between the edge channels along an array of tunnel junctions. The line junction is assumed to have a sufficiently large length $L$ compared with the equilibration length $\ell_{\rm eq}$ so that the edge channels and the magnons equilibrate fully. The equilibration continues until the net tunneling between the edge channels vanishes. While, in principle, one can imagine a situation in which the edge modes in this regime would be characterized by a non-trivial out-of-equilibrium distribution, we assume here that inelastic mechanisms allow the full system (the edge channels plus magnons) to relax to an equilibrium distribution. In the following, we use this line junction model to calculate the electrical current and noise generated in the magnon absorption regions.

We begin with the magnon generation process. As described above, the edge channels achieve full equilibration at positions $x \gg \ell_{\rm eq}$, where the net tunneling current vanishes, i.e., $\langle I_{\rm ge}(x) \rangle = 0$. We thus set

$$\langle I_{\rm ge}(x) \rangle = \frac{2\pi}{\hbar} e |W_{\rm ge}|^2 \rho_{\rm edge}^2 \rho_m \int d\omega_1 \int d\omega_2 \theta(\omega_2 - \omega_1 - E_Z)$$
$$\times \Big[ (1 - f_\uparrow(\omega_1, x)) f_\downarrow(\omega_2, x)(1 + b(\omega_2 - \omega_1 - \mu_m))$$
$$- f_\uparrow(\omega_1, x)(1 - f_\downarrow(\omega_2, x)) b(\omega_2 - \omega_1 - \mu_m) \Big] = 0. \tag{S33}$$

Here, we assumed that the magnons and the edge channels have relaxed to individual equilibrium distribution functions $f_s$ and $b$ and used the corresponding generalization of Eq. (S14). The simplest solution for distribution functions that satisfy Eq. (S33) is that the edge channels and the magnon system acquire the same temperature $T$ ("true equilibration"), i.e.,

$$f_{\uparrow/\downarrow}(\omega) = \frac{1}{e^{(\omega - \mu_{\uparrow/\downarrow}^{\rm ge})/T} + 1}, \quad b(\omega) = \frac{1}{e^{(\omega - \mu_m)/T} - 1}, \tag{S34}$$

$$\mu_\downarrow^{\rm ge} - \mu_\uparrow^{\rm ge} = \mu_m. \tag{S35}$$

However, for a low temperature and a reasonable size of the line junction $L$, this true equilbration will not be reached. Indeed, the equilibration process becomes strongly suppressed once the chemical potential difference between the edge channels reduces down to $E_Z$,

$$\mu_\downarrow^{\rm ge} - \mu_\uparrow^{\rm ge} = E_Z. \tag{S36}$$

23

A further magnon generation would correspond to $\mu_\downarrow^{\text{ge}} - \mu_\uparrow^{\text{ge}}$ becoming smaller than $E_Z$ and is strongly suppressed with an exponential factor $e^{(\mu_\downarrow^{\text{ge}} - \mu_\uparrow^{\text{ge}} - E_Z)/T}$. Therefore, for a realistic (not exponentially large) length $L$ of the line junction, the equilibration effectively stops at the saturation point (S36). While this is a kind of "pre-equilibration", we term this situation "full equilibration" for brevity. Correspondingly, the case of a short junction, when resulting electrochemical potentials do not reach Eq. (S36), i.e., $\mu_\downarrow^{\text{ge}} - \mu_\uparrow^{\text{ge}} > E_Z$ is termed in this paper "partial equilibration".

From charge conservation, we can relate the electrochemical potential of the edge channels emanating from the source contact with $\mu_{\uparrow/\downarrow}^{\text{ge}}$ as

$$\mu_\downarrow^{\text{ge}} + \mu_\uparrow^{\text{ge}} = |eV_S|. \tag{S37}$$

Similarly, we obtain from charge conservation an equation for the absorption regions

$$\mu_\downarrow^{\text{ab},i} + \mu_\uparrow^{\text{ab},i} = 0. \tag{S38}$$

Here, $\mu_{\downarrow,\uparrow}^{\text{ab},i}$ are the electrochemical potentials after the magnon absorption process in each individual magnon-absorption region. The electrochemical potentials $\mu_{\uparrow,\downarrow}^{\text{ab},i}$ in the absorption region are related to the electrochemical potentials $\mu_{\uparrow,\downarrow}^{\text{ge}}$ in the generation region by the steady state condition

$$I_{\text{ge}} = -\sum_{i=1}^{M} I_{\text{ab}}^i, \tag{S39}$$

where $M$ is the number of absorption regions. Equation (S39) guarantees that in the steady state the magnon generation and absorption rates are identical, keeping the number of magnons in the bulk constant. Combining Eqs. (S36)-(S39), we thus obtain

$$\mu_\downarrow^{\text{ge}} = \frac{|eV_S| + E_Z}{2}, \quad \mu_\uparrow^{\text{ge}} = \frac{|eV_S| - E_Z}{2}, \quad \mu_\downarrow^{\text{ab}} = \frac{|eV_S| - E_Z}{2M}, \quad \mu_\uparrow^{\text{ab}} = -\frac{|eV_S| - E_Z}{2M}. \tag{S40}$$

Here, we assumed for simplicity that in each individual absoprion region, the magnons are absorbed with equal probabilities, and hence $\mu_\downarrow^{\text{ab},i} = \mu_\downarrow^{\text{ab}}$ and $\mu_\uparrow^{\text{ab},i} = \mu_\uparrow^{\text{ab}}$. We further find an equation for energy conservation, which reads for our experimental geometry (shown in Fig. 1(a) in the main text)

$$2 \times \frac{(eV_S)^2}{2h} + 6 \times \frac{\pi^2 T_0^2}{6h}$$
$$= \frac{(|eV_S| + 2\mu_\uparrow^{\text{ab}})^2 + (2\mu_\downarrow^{\text{ab}})^2 + \sum_{s=\uparrow,\downarrow}((\mu_s^{\text{ge}} + \mu_s^{\text{ab}})^2 + (2\mu_s^{\text{ab}})^2)}{2h} + 6 \times \frac{\pi^2 T^2}{6h}. \tag{S41}$$

Here, we used the fact that, in our geometry, there are in total six edge channels and out of them two channels are biased. Further, $T_0$ is the ambient temperature of the contacts and $T$ is the effective temperature of the system as a result of equilibration. We assume that the edge modes reach equilibrium with the entire system sharing the same temperature $T$. We extract this temperature from Eq. (S40) and (S41) and it reads

$$T = \sqrt{T_0^2 + \frac{3(|eV_S| - E_Z)(2E_Z + 3|eV_S|)}{(5\pi)^2}\theta(|eV_S| - E_Z)}. \tag{S42}$$

As derived in the next subsection, the total noise in the individual absorption region is given by

$$S^i = \frac{e^2}{h} k_B (T + T_0) \quad (S43)$$

and thus the corresponding excess noise is given by

$$S_I^i = \frac{e^2}{h} k_B (T_0 + T) - 2\frac{e^2}{h} k_B T_0 = \frac{e^2}{h} k_B (T - T_0). \quad (S44)$$

The superscript $i$ denotes the five absorption regions. The tunneling current and the corresponding differential non-local conductance at an absorption region are given by

$$I_{\text{ab}} = \frac{e}{2Mh} \theta(|eV_S| - E_Z)(|eV_S| - E_Z), \quad (S45)$$

$$\frac{dI_{\text{ab}}}{dV} = \frac{e^2}{2Mh} \theta(|eV_S| - E_Z). \quad (S46)$$

Note that temperature corrections to the dc current are exponentially small (cf. Eq. (S36) and the following discussion). By contrast, the noise is linearly dependent on the effective temperature.

We can define an effective Fano factor in the individual absorption region as $F_{\text{ab}} \equiv S_I/(2eI_{\text{ab}})$. Using that $M = 5$ in our geometry, we find that, in the strong bias regime ($|eV_S| \gg E_Z, T_0$), the Fano factor approaches

$$F_{\text{ab}} \to \frac{3}{\pi} \approx 0.95. \quad (S47)$$

The measured current and the excess noise are given by

$$I = \frac{\langle I_{\text{ab}}^{\mathbf{B}} \rangle - \langle I_{\text{ab}}^{\mathbf{D}} \rangle}{2}, \quad S_I = \frac{S_I^{\mathbf{B}} + S_I^{\mathbf{D}}}{4}. \quad (S48)$$

Importantly, the factor of 1/2 entering the current and the factor 1/4 for the excess noise come from the experimental geometry for measuring these quantities, see Fig. 1(a) in the main text and Supplementary Fig. 6(a). More specifically, the floating contact in these figures emits currents and fluctuations on two edges of the device. The contact measuring the floating-contact electrochemical potential $\mu_{\text{FC}}$ and its fluctuations $\delta\mu_{\text{FC}}$ is located on one of these edges and thus receives only half of the emission. We have also taken into account that two absorption regions '**B**' and '**D**' contribute to the measured noise and the measured current.

The non-local conductance and the noise in the line junction model display $eV_S$-characteristics distinct from those in the single tunnel junction model. While the non-local conductance in the single tunnel junction is linear in $eV_S$, the line junction model has a constant differential conductance, see Eq. (S46). Further, the noise at large bias increases quadratically with $eV_S$ in the single tunnel junction, whereas it grows linearly as a function of $eV_S$ in the line junction model, see Eqs. (S44) and (S42). We plot the calculated excess noise vs $eV_S$ at a few different temperatures in Supplementary Fig. 10(a). At low temperatures,

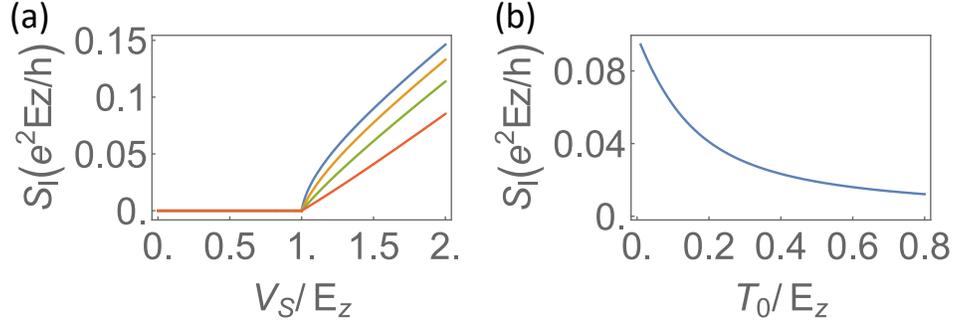

**Supplementary Figure 10: Excess noise in the line junction model** (a) as function of the bias energy $eV_S$ at different temperatures (blue: $T_0 = 0.02E_Z$, orange: $T_0 = 0.05E_Z$, green: $T_0 = 0.1E_Z$, red: $T_0 = 0.2E_Z$) and (b) as a function of temperature $T_0$ at fixed bias voltage $eV_S = 1.5E_Z$. We set $e = k_B = 1$ for these plots.

$T_0 \to 0$, we find $S_I \propto \sqrt{|eV_S| - E_Z}\theta(|eV_S| - E_Z)$ in the vicinity of $|eV_S| = E_Z$. The cusp-like singularity at $|eV_S| = E_Z$ is an artifact of the full-equilibration assumption, which, for fixed junction length $L$, in fact ceases to be applicable when $|eV_S|$ approaches $E_Z$, see next subsection. We also plot the temperature dependence of the excess noise at fixed voltage $eV_S$ in Supplementary Fig. 10(b). We see that the excess noise decreases with increasing temperature, which is in full consistency with the experimental observation presented in Fig. 3(e) in the main text.

**S9.5. Noise generated in a line junction.** In this subsection, we present a derivation of the formula (S43) (equivalently, Eq. (1) in the main text) for the noise of the tunneling current generated in the magnon absorption regions. A scheme of the line junction formed by two co-propagating edges in an absorption region is shown in Supplementary Fig. 11. In such a region, electron tunneling between the two edge channels is only possible by absorption of impinging magnons, in view of the angular momentum conservation constraint. We assume that the line junction consists of a series of tunnel junctions. Under the assumption of full equilibration of the edge channels, we show that the noise is dominantly generated at the downstream end of the line junction (indicated by the yellow circle in Fig. 11). The magnitude of noise is governed by the effective temperature in this "noise spot". This noise-generating mechanism, which is described in more detail below, is essentially the same as that identified and analyzed in Refs. [10] and [11] in the context of fractional quantum Hall edges.

We begin by investigating a small segment of the line junction which contains only a single tunnel junction. Charge conservation relates the two channel charge currents as

$$I_{j,\uparrow} = I_{j-1,\uparrow} - I_{\text{tun},j}, \quad I_{j,\downarrow} = I_{j-1,\downarrow} + I_{\text{tun},j}. \tag{S49}$$

Here, $I_{j,\uparrow/\downarrow}$ is the charge current flowing along segment $j$ ($1 \leq j \leq N$) of the edge channel with spin $\uparrow/\downarrow$,

26

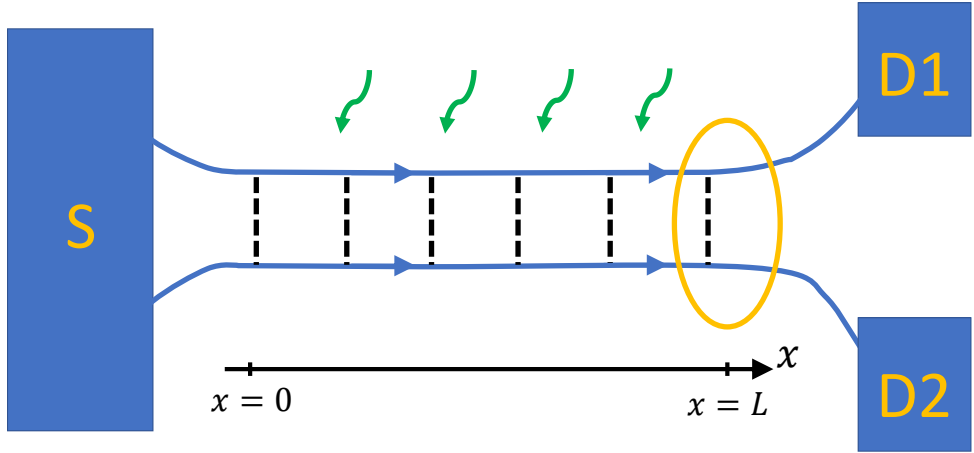

**Supplementary Figure 11: Schematic plot of a line junction** in a magnon-absorption region. The electron tunneling between two co-propagating edge modes is allowed by absorbing magnons (indicated by green arrows). The noise of the total tunneling current is mostly generated in the vicinity of $x = L$ ("noise spot") shown by a yellow circle).

$I_{\text{tun},j}$ represents the tunneling current in tunnel junction $j$, and the fluctuations in $I_{\text{tun},j}$ is decomposed as

$$\delta I_{\text{tun},j} = g_j \frac{e^2}{h}(\delta V_{j-1,\uparrow} - \delta V_{j-1,\downarrow}) + \delta I^{\text{int}}_{\text{tun},j} = g_j(\delta I_{j-1,\uparrow} - \delta I_{j-1,\downarrow}) + \delta I^{\text{int}}_{\text{tun},j}. \tag{S50}$$

The first contribution in Eq. (S50) comes from the fluctuations currents impinging (from the left) on the tunnel junction. The second contribution describes intrinsic current fluctuation generated in the junction. This term includes contributions of local-equilibrium (i.e., thermal) and out-of-equilibrium (i.e., shot-noise) origin. Finally, $g_j$ is the tunneling probability, which generically depends on the strength of the local tunneling current and how the incoming states occupy microscopic states. We assume $g_j$ to be weak: $g_j \ll 1$. By using Eqs. (S49) and (S50), we can relate the currents in each segment according to

$$\begin{pmatrix} \delta I_{j,\uparrow} \\ \delta I_{j,\downarrow} \end{pmatrix} = M_j \begin{pmatrix} \delta I_{j-1,\uparrow} \\ \delta I_{j-1,\downarrow} \end{pmatrix} + \boldsymbol{v}_j, \tag{S51}$$

with the matrix $M_j$ and the vector $\boldsymbol{v}_j$ are given by

$$M_j = \begin{pmatrix} 1-g_j & g_j \\ g_j & 1-g_j \end{pmatrix}, \quad \boldsymbol{v}_j = \delta I^{\text{int}}_{\text{tun},j} \begin{pmatrix} -1 \\ 1 \end{pmatrix}. \tag{S52}$$

Equation (S51) can be solved recursively with the solution

$$\begin{pmatrix} \delta I_{N,\uparrow} \\ \delta I_{N,\downarrow} \end{pmatrix} = \left(\prod_{j=1}^{N} M_{N-j+1}\right) \begin{pmatrix} \delta I_{0,\uparrow} \\ \delta I_{0,\downarrow} \end{pmatrix} + \sum_{j=1}^{N} \left(\prod_{j'=1}^{N-j} M_{N-j'+1}\right) \boldsymbol{v}_j. \tag{S53}$$

In the limit $\sum_{j=1}^{N} g_j \gg 1$, we can neglect term exponentially suppressed in $N$ so that Eq. (S53) reduces to

$$\begin{pmatrix} \delta I_{N,\uparrow} \\ \delta I_{N,\downarrow} \end{pmatrix} \approx \frac{\delta I_\uparrow^0 + \delta I_\downarrow^0}{2} \begin{pmatrix} 1 \\ 1 \end{pmatrix} - \sum_{j=1}^{N} e^{-2\sum_{j'=j+1}^{N} g_{j'}} \delta I_{\text{tun},j}^{\text{int}} \begin{pmatrix} 1 \\ -1 \end{pmatrix}. \tag{S54}$$

Next, we assume uniform tunneling, i.e., $g_j = g$, and take the continuum limit (justified by large $N$ and small $g$). We then obtain the noise measured in drains $D1$ and $D2$ as

$$S_{D1} = S_{D2} = \overline{(\delta I_{N,\uparrow})^2} = \overline{(\delta I_{N,\downarrow})^2} = \frac{e^2}{h} T_0 + \frac{1}{a} \int_0^L dx\, e^{-4g\frac{L-x}{a}} \overline{(\delta I_{\text{tun}}^{\text{int}}(x))^2}. \tag{S55}$$

Here, $\overline{(\delta I_{N,s})^2} \equiv \int dt (\overline{\delta I_{N,s}(t) \delta I_{N,s}(0)} + \overline{\delta I_{N,s}(0) \delta I_{N,s}(t)})$ denotes the zero frequency noise of $\delta I_{N,s}(t)$ with $s = \uparrow, \downarrow$, $a$ is the distance between two consecutive tunnel junctions, and $L$ is the size of the line junction, i.e., $L = Na$. In Eq. (S55), we have used the fact that the intrinsic tunneling currents from different junctions are not correlated to each other. Note also that the exponential factor in the integrand shows explicitly that the noise is dominantly generated in the vicinity of $x = L$ ("noise spot") in the equilibrated regime, $L \gg \ell_{\text{eq}}$, where $\ell_{\text{eq}} = a/4g$ is the equilibration length. In the noise spot, the two edge modes will be fully equilibrated such that $\mu_m = \mu_\downarrow - \mu_\uparrow$, and thus the equilibrium noise dominates over the shot noise contribution to $\overline{(\delta I_{\text{tun}}^{\text{int}}(x))^2}$. We may therefore approximate $\overline{(\delta I_{\text{tun}}^{\text{int}}(x))^2}$ as

$$\overline{(\delta I_{\text{tun}}^{\text{int}}(x))^2} \approx 4g \frac{e^2}{h} k_B T, \tag{S56}$$

where $T$ is the local temperature of the edge channels after equilibration. By inserting the local noise (S56) into Eq. (S55) and performing the integral, we arrive in the limit of full equilibration, $L/\ell_{\text{eq}} = 4Lg/a \gg 1$, at our final expression

$$S_{D1} = S_{D2} = \frac{e^2}{h} k_B (T + T_0). \tag{S57}$$

This completes the derivation of Eq. (S43) of the preceding section.

**S9.6. Voltage dependence of equilibration length, overall behavior of the noise, and comparison to experimental data.** In this subsection, we calculate the dependence of the equilibration length on the bias voltage $|eV_S|$. This allows us to demonstrate that the single tunnel junction model is applicable for $|eV_S|$ sufficiently close to $E_Z$, while the line junction model works for larger $|eV_S|$. We then compare the theoretical predictions with the experimental data.

To investigate the bias voltage dependence of the equilibration length $\ell_{\text{eq}}$ in the magnon generation region, we consider a series of $N'$ tunnel junctions, neglecting possible quantum-interference effects between them. The tunneling currents in the individual tunnel junction are then added up and contribute the total tunnel current as $I_{\text{ge}}^{\text{tot}} = \sum_{j=1}^{N'} I_{\text{ge}}^j$. By using the expression for the tunneling current at a single tunnel junction, Eq. (S29), we arrive at

$$I_{\text{ge}}^{\text{tot}} = \frac{e}{2h} N' \gamma^{\text{ge}} (|eV_S| - E_Z)^2 \theta(|eV_S| - E_Z), \tag{S58}$$

where we assumed that $\gamma^{\text{ge}}$ is the same for each tunnel junction. This result is valid as long as $N'$ is not too large, so that the line junction effectively works as a tunnel junction with $\gamma^{\text{ge}} \to N'\gamma^{\text{ge}}$. The equilibration length is thus obtained as $\ell_{\text{eq}} = N'a$ (with $a$ being the distance between the adjacent junctions) at which $I_{\text{ge}}^{\text{tot}}$ is equal to the total tunneling current $\frac{e}{2h}(|eV_S| - E_Z)$ by the full equilibration, i.e.,

$$I_{\text{ge}}^{\text{tot}} = \frac{e}{2h}N'\gamma^{\text{ge}}(|eV_S| - E_Z)^2\theta(|eV_S| - E_Z) = \frac{e}{2h}(|eV_S| - E_Z)\theta(|eV_S| - E_Z). \quad \text{(S59)}$$

From this equation, we obtain an expression for $\ell_{\text{eq}}$ as

$$\ell_{\text{eq}} = N'a = \frac{a}{\gamma^{\text{ge}}(|eV_S| - E_Z)} \equiv \frac{1}{\gamma(|eV_S| - E_Z)}. \quad \text{(S60)}$$

Here, $\gamma \equiv \gamma^{\text{ge}}/a$ is a parameter associated with the tunneling strength in the tunnel junctions. When the voltage $|eV_S| > E_Z$ approaches $E_Z$, the equilibration length diverges and thus prevents the edge channels from reaching the full equilibration. Thus, there is an intermediate regime of $|eV_S|$ slightly exceeding $E_Z$ that can not be captured by full equilibration of the line junction model considered in the previous subsection, but is rather effectively described by the single tunnel junction model. This leads in particular to a smearing of the cusp-like behavior of noise at $|eV_S| = E_Z$ obtained in the line junction model.

Based on the bias voltage dependence of $\ell_{\text{eq}}$ calculated above [see Eq. (S60)], we thus identify the following three regimes, see Supplementary Figure 12: (i) Biases $|eV_S| < E_Z$ result in no magnon generation and thus no excess noise (discarding exponentially small contributions). (ii) In a narrow region $0 < |eV_S| - E_Z < \frac{1}{\gamma L}$ (which corresponds to $\ell_{\text{eq}} \equiv \frac{1}{\gamma(|eV_S| - E_Z)} > L$), the equilibration in magnon absorption and generation regions is only partial, as in the single tunnel junction model. In this regime, the noise generation is of non-equilibrium nature, resulting in $S_I \propto (|eV_S| - E_Z)^2$, see Eq. (S32). (iii) For larger biases $|eV_S| > E_Z + \frac{1}{\gamma L}$ and hence $\ell_{\text{eq}} < L$, the edge channels and magnons achieve full equilibration in the magnon absorption and generation regions.

In Supplementary Figure 12, these theoretical predictions are compared to typical experimental data. The green and red line show the theoretical result for the single tunnel junction and line junction models that are applicable in the regimes (ii) and (iii), respectively. These predictions are in a good agreement with experimental observations. Furthermore, we find that for $|eV_S|$ sufficiently exceeding $E_Z$ [regime (iii)], the line junction model correctly reproduces several experimental observations: (1) a rather sharp increase, followed by a saturation, of the non-local conductance as a function of the bias voltage [compare Fig. 1(f) in the main text with Eq. (S46)], (2) a linear behavior of the noise as a function of the bias voltage [compare Fig. 2(d) with Supplementary Fig. 10(a)], and (3) the temperature dependence of the excess noise [compare Fig. 3(e) with Supplementary Fig. 10(b)].

**References**

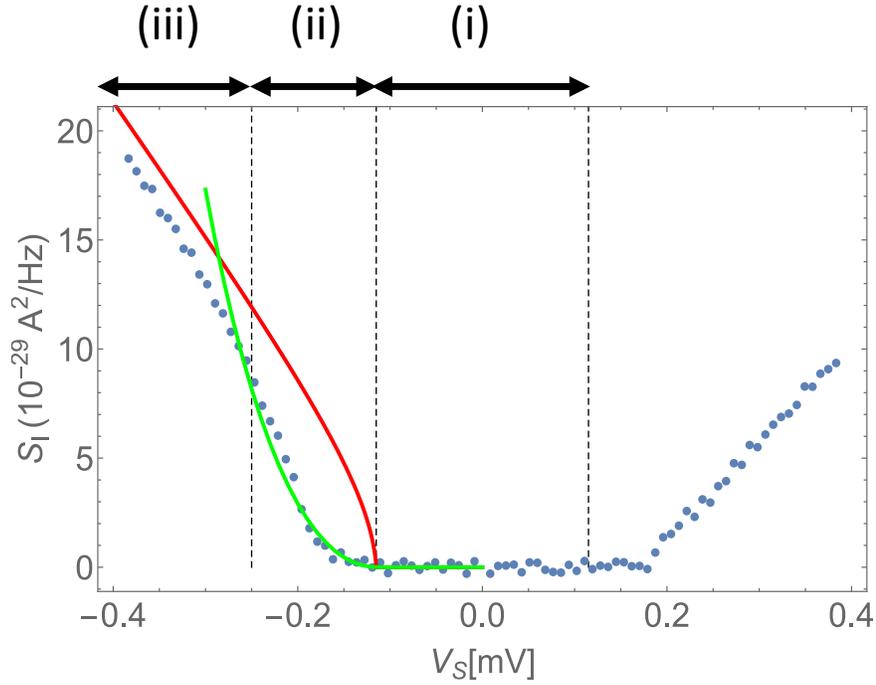

**Supplementary Figure 12: Three different regimes** of the noise $S$ vs the bias voltage $V_S$: (i) regime of no magnon generation, (ii) partially equilibrated regime, and (iii) fully equilibrated regime. While the single tunnel junction model (green line) describes regime (ii), regime (iii) is captured by the line junction model (red line). The blue dots represent experimental data (the same set as in Fig. 3(b) in the main text). The value of $\gamma$ in regime (ii) is chosen from a fit of the experimental data to Eqs. (S58) and (S60), which yields a dimensionless parameter $\gamma E_Z L = 0.8$.



\end{document}